\documentclass{emulateapj}
\newcommand{\tens} [1] {\overline{\overline{\bf #1}}}

\shorttitle{Small Scale turbulence in the Solar Wind}
\shortauthors{Sahraoui et al.}

\begin{document}


\title{NEW INSIGHT INTO SHORT WAVELENGTH SOLAR WIND FLUCTUATIONS FROM VLASOV THEORY}


\author{F. Sahraoui\altaffilmark{1} and G. Belmont\altaffilmark{1}}
\affil{Laboratoire de Physique des Plasmas, CNRS-Ecole Polytechnique-UPMC, Observatoire de Saint-Maur, 4 avenue de Neptune, 94107 Saint-Maur-des-Foss\'es, France}

\and

\author{M. L. Goldstein\altaffilmark{2}}
\affil{NASA Goddard Space Flight Center, Code 673, Greenbelt 20771, Maryland, USA}
\email{fouad.sahraoui@lpp.polytechnique.fr}

%\altaffiltext{1}{Visiting Astronomer, Cerro Tololo Inter-American Observatory.
%CTIO is operated by AURA, Inc.\ under contract to the National Science
%\altaffiltext{2}{Society of Fellows, Harvard University.}

\begin{abstract}
The nature of solar wind (SW) turbulence below the proton gyroscale is a topic that is being investigated extensively nowadays, both theoretically and observationally. Although recent observations gave evidence of the dominance of Kinetic Alfv\'en Waves (KAW) at sub-ion scales with $\omega<{\omega_{ci}}$, other studies suggest that the KAW mode cannot carry the turbulence cascade down to electron scales and that the whistler mode (i.e., $\omega>\omega_{ci}$) is more relevant. Here, we study key properties of the short wavelength plasma modes under limited, but realistic, SW conditions, typically $\beta_i\gtrsim \beta_e\sim 1$ and for high oblique angles of propagation $80^\circ\leq \Theta_{\bf kB}<90^\circ$ as observed from the Cluster spacecraft data. The linear properties of the plasma modes under these conditions are poorly known, which contrasts with the well-documented cold plasma limit and/or moderate oblique angles of propagation ($\Theta_{\bf kB} <80^\circ$). Based on linear solutions of the Vlasov kinetic theory, we discuss the relevance of each plasma mode (fast, Bernstein, KAW, whistler) in carrying the energy cascade down to electron scales. We show, in particular, that the shear Alfv\'en mode (known in the MHD limit) extends at scales $k\rho_i\gtrsim1$ to frequencies either larger or smaller than $\omega_{ci}$, depending on the anisotropy $k_\parallel/ k_\perp$. This extension into small scales is more readily called whistler ($\omega>\omega_{ci}$) or KAW ($\omega<\omega_{ci}$), although the mode is essentially the same. This contrasts with the well-accepted idea that the whistler branch develops always as a continuation at high frequencies of the fast magnetosonic mode. We show, furthermore, that the whistler branch is more damped than the KAW one, which makes the latter the more relevant candidate to carry the energy cascade down to electron scales. We discuss how these new findings may facilitate resolution of the controversy concerning the nature of the small scale turbulence, and we discuss the implications for present and future spacecraft wave measurements in the SW.
\end{abstract}

\keywords{turbulence, solar wind, heating, whistler, KAW}

%--------------------------------------------------------------------------------------------
% New section 
%--------------------------------------------------------------------------------------------
\section{Introduction}
The nature of the SW turbulence below the ion scale (typically $\rho_i \sim 100$ km, corresponding to an observed frequency in the spacecraft frame of $f_{sc} \sim 0.5$ Hz) has attracted considerable interest in the space and astrophysical communities in recent years. This has been encouraged in particular by recent observations from the Cluster mission that provided the most complete and detailed picture of the SW turbulence cascade from magnetohydrodynamic (MHD) scales ($L>>\rho_i$) to electron scales ($L\sim\rho_e\sim 1$ km)~\citep{sahraoui09,kiyani09,alexandrova09,chen10,sahraoui10a}. Determining the nature and properties (e.g., scaling, anisotropy) of the turbulence at small scales is indeed a crucial point to understanding the problems of energy dissipation and heating, particle acceleration, and magnetic reconnection  in space and astrophysical plasmas \citep{schekochihin09}. Recent Cluster observations provided clear evidence that SW turbulence cascades below the ion scale $\rho_i$ down to the electron scale $\rho_e$ where dissipation becomes important and the spectra steepen to $\sim {f_{sc}}^{-\alpha}$, with $\alpha \gtrsim 4$~\citep{sahraoui09,sahraoui10a}. The spectrum thus formed was termed the {\it dissipation range} while the range of scales between $\rho_i$ and $\rho_e$ has been termed the {\it dispersive range} in reference to the dispersive nature of the plasma modes at those scales~\citep{stawicki01}. In a case study, \cite{sahraoui10a} performed a detailed analysis of the energy cascade from MHD to sub-ion scales 
using a multipoint measurement technique called {\it k}-filtering~\citep{pincon91,sahraoui03a, sahraoui10b, narita10a, narita10b, tjulin05}. The results showed clearly that: i) The magnetic turbulence is strongly anisotropic ($k_\perp>>k_\parallel$) down to the observed scale $k_\perp\rho_i\sim 2$; ii) The cascade is consistent with KAW turbulence as proposed in~\cite{howes08a} and~\cite{schekochihin09} with frequencies in the plasma rest frame $\omega \lesssim0.1\omega_{ci}$ although the frequencies in the spacecraft frame reached $20\omega_{ci}$; iii) The turbulence undergoes a {\it transition range} near the ion scale $\rho_i$ characterized by a steepening of the spectrum from $ k_\perp^{-1.6}$ to $k_\perp^{-4.5}$, which has been interpreted as due to Landau damping of magnetic energy into ion heating~\citep{howes08b}. The remaining energy cascades following a power-law $\sim f_{sc}^{-2.8}$ down to the electron scale where the energy was suggested to dissipate into electron heating. The wave number spectra at those frequencies could not have been {\it directly} measured using the {\it k}-filtering technique due to the limitation imposed by the satellites separations, which are larger than $100$ km~\citep
{sahraoui10c}. However, 1D wavenumber spectra can be inferred indirectly by means of the Taylor frozen-in-flow approximation provided that it is valid~\citep{bale05, sahraoui09, alexandrova09}. Similar analyses of magnetic turbulence at MHD and sub-ion scales using the {\it k}-filtering technique have been carried out recently and have confirmed most of these results~\citep{narita11,sahraoui11, narita10a, narita10b}\footnote{\cite{narita11} have claimed to have observed several plasma modes, other than the KAW, based on the random spread of the observed dispersion relations. However, given the absence of any statistical error bars in those observations, one cannot reliably interpret those results. A detailed discussion of these important issues will be given elsewhere.}

From the theoretical point of view, the debate is generally polarized between those who advocate for the relevance of Kinetic Alfv\'en Wave (KAW) turbulence to explain observations of small scale SW turbulence~\citep{howes08a, howes08b, schekochihin09,howes11a} and those who believe in the necessity that whistler (or another type) of turbulence should exist at small scales~\citep{stawicki01,gary09,podesta10}. Indeed, \cite{podesta10} have used a toy model of energy cascade where the linear damping rate of the KAW waves is considered and showed that KAW cascade cannot reach the electron scale, because the energy flux of the cascade vanishes at scales $k\rho_i\sim 20$. However, recent GyroKinetic (GK) simulations, which self-consistently contain kinetic damping of the plasma modes, did not confirm that conclusion and showed rather that the KAW mode can carry turbulence cascade down to electron scales~\citep{howes11a}. Recent 2D PIC simulations showed also the formation of a power law spectrum $k_\perp^{-5.8}$ below the electron inertial length~\citep{camporeale11}. In the framework of incompressible Electron-MHD turbulence \cite{mayrand10} showed that  the magnetic energy spectra should follow a power law $k^{-11/3}$ at scales smaller than $d_e$ (the electron inertial length), which they proposed to explain the steepening of the spectra near the electron scale reported in \cite{sahraoui09,sahraoui10a}. Note however that this fluid model is non-dissipative and does not consider any damping of the turbulence via kinetic effects, which are important in the dispersive and the dissipation ranges.

Here we derive and analyze the linear solutions of the Vlasov-Maxwell equations to shed light on new aspects of small scale kinetic plasma modes.  Based on the obtained damping rates of the waves and their magnetic compressibilities we discuss the relevance of each mode to carry the energy cascade in the SW down to electron scales. Unlike the previous papers that had dealt with this problem, in this article we focus only on the properties of the plasma modes under a restricted range of SW parameters, namely hot plasma with $\beta_i \gtrsim\beta_e \sim 1$ and high oblique angles of propagation, typically $80^\circ\leq\theta_{\bf kB}<90^\circ$, as observed recently using the {\it k}-filtering technique~\citep{sahraoui10a,narita11,sahraoui11}. This range of parameters has not been addressed carefully so far. Filling this gap should help resolving part of the ongoing controversies on the problem of energy cascade and dissipation in the dispersive range. 

Before studying in detail the plasma kinetic modes, in the next section we will first derive  the linear solutions of the two-fluid theory in the cold and hot plasma cases. Although one may question the validity of the fluid description of hot plasmas (we do not address that issue here, see~\cite{krauss94,howes09}),  we believe nevertheless that deriving the fluid solutions is a necessary step to better understand the more complex solutions given by the kinetic theory. This complexity is essentially due to the damping of the modes by kinetic effects, which makes them hard to trace at all scales, in addition to the rising of several mode conversions~\citep{krauss94,li01}. As we will show, the reduced-two-fluid theory indeed helps tracking the various plasma modes and understanding the important changes in their properties depending on the plasma $\beta$ and the obliquity of the waves. It allows us also to derive analytically various asymptotic properties of the waves that will be used to discuss new properties of the Alfv\'en and the whistler modes. In the last section we discuss the results and their implications on spacecraft measurements of waves in the SW. 

%--------------------------------------------------------------------------------------------
% New section 
%--------------------------------------------------------------------------------------------
\section{Linear solutions of the reduced-two-fluid model}
We use the reduced-two-fluid theory as described in~\cite{sahraoui03b}, whose equations are the following
\begin{eqnarray}
\label{eq1}
\partial_tn_s+\nabla.(n_s{\bf v}_s) &=&0 \\
\label{eq2}
\partial_t{\bf v}_s+{\bf v}_s.\nabla{\bf v}_s&=& -\frac{\nabla P_s}{m_sn_s}+\frac{q_s}{m_s}({\bf E}+{\bf v}_s\times{\bf B}) \\
\label{eq3}
d_t\Bigg[\frac{P_s}{(m_sn_s)^{\gamma_s}}\Bigg]&=&0 \\
\label{eq4}
\qquad \partial_t{\bf B}&=&-\nabla\times{\bf E}\\ 
\label{eq5}
\nabla\times{\bf B}&=&\mu_0{\bf J}
\end{eqnarray}
where the subscript $s$ denotes ions and electrons, $\gamma_s$ is the polytropic index and ${\bf J}=\sum_sq_sn_s{\bf v}_s$ is the electric current. The system of equations~(\ref{eq1}-\ref{eq5}) is essentially equivalent to the classical two-fluid model when the non-relativistic and quasi-neutrality assumptions are considered. This translates into neglecting both the longitudinal and the transverse components of the displacement current, which yields the simplified form of equation~\ref{eq5}. This system has the advantage of ruling out the three high frequency modes of the two-fluid theory: the optic modes (with $V_\phi\sim c$, the speed of light) and the Langmuir mode $\omega\sim\omega_{pe}$. However, it still includes the electron inertia, which allows us to describe properly the whistler mode. The system has therefore three eigenmodes whose dispersion relations can be derived analytically. The main simplification with respect to the complete kinetic system that will be studied in the next section concerns the introduction of polytropic closure equations that excludes any resonant phenomenon, in particular any effect of wave damping. 

\begin{figure}
\includegraphics[height=7cm,width=8cm]{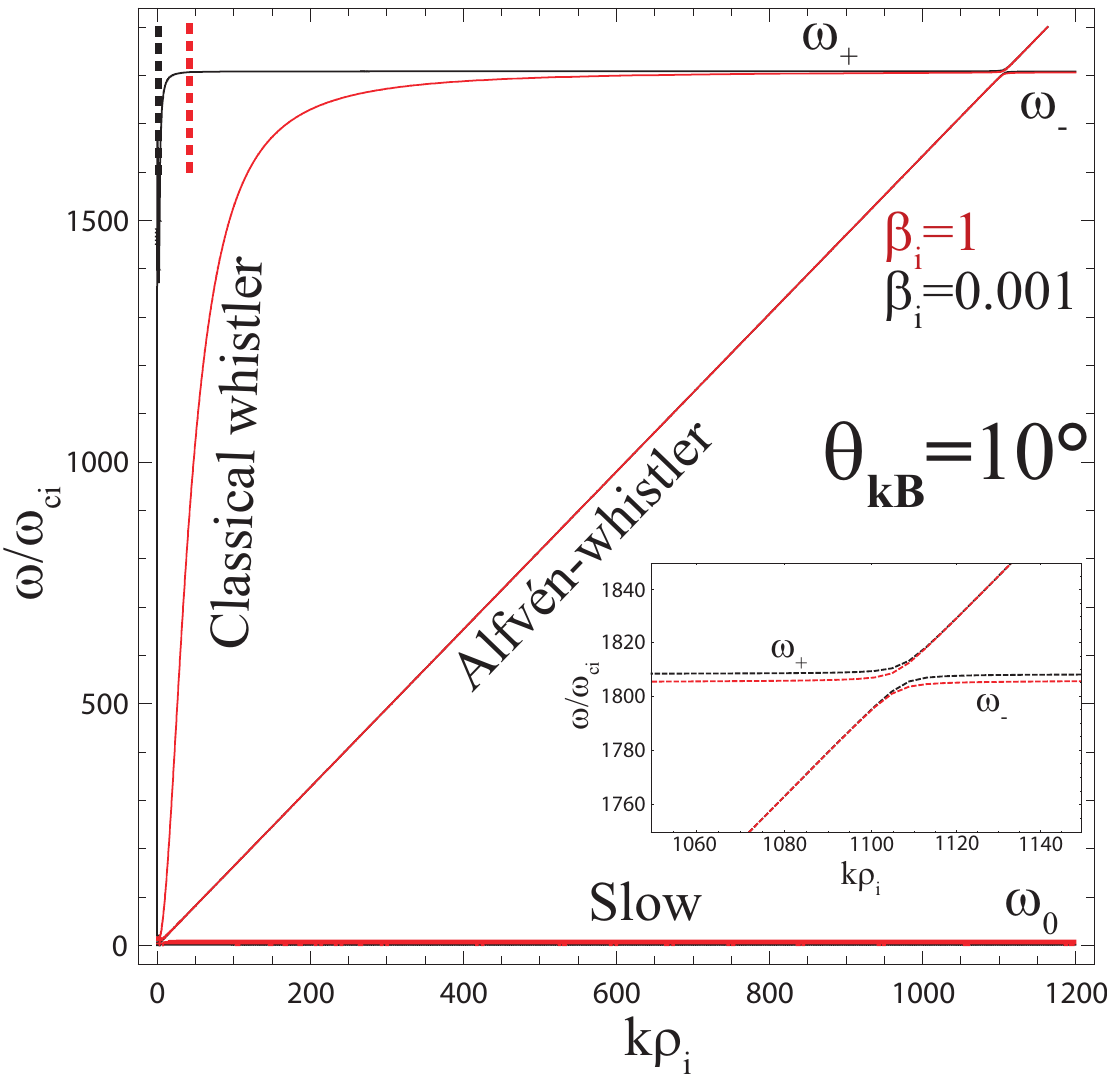}
\includegraphics[height=7cm,width=8cm]{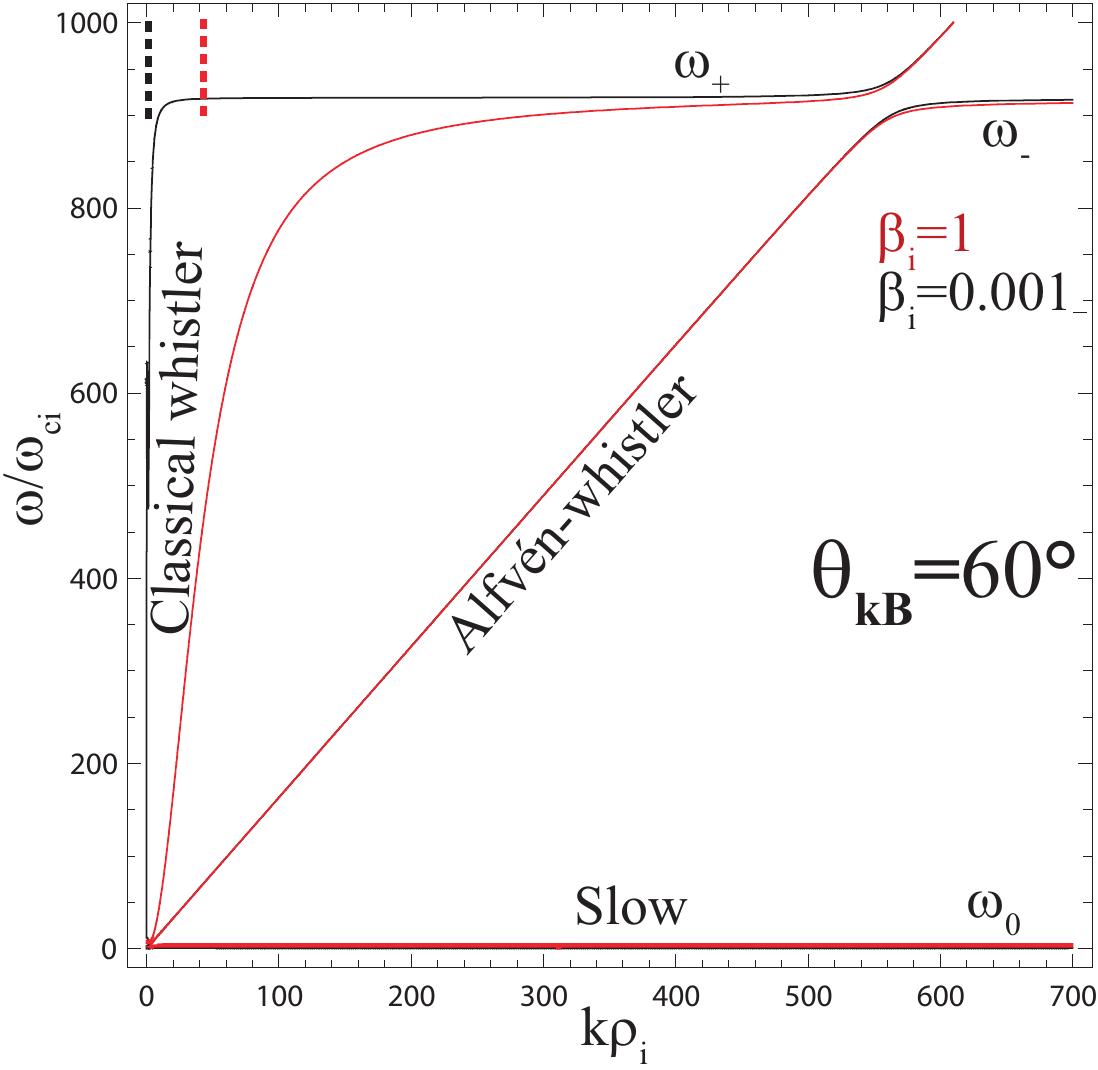}
\includegraphics[height=7cm,width=8cm]{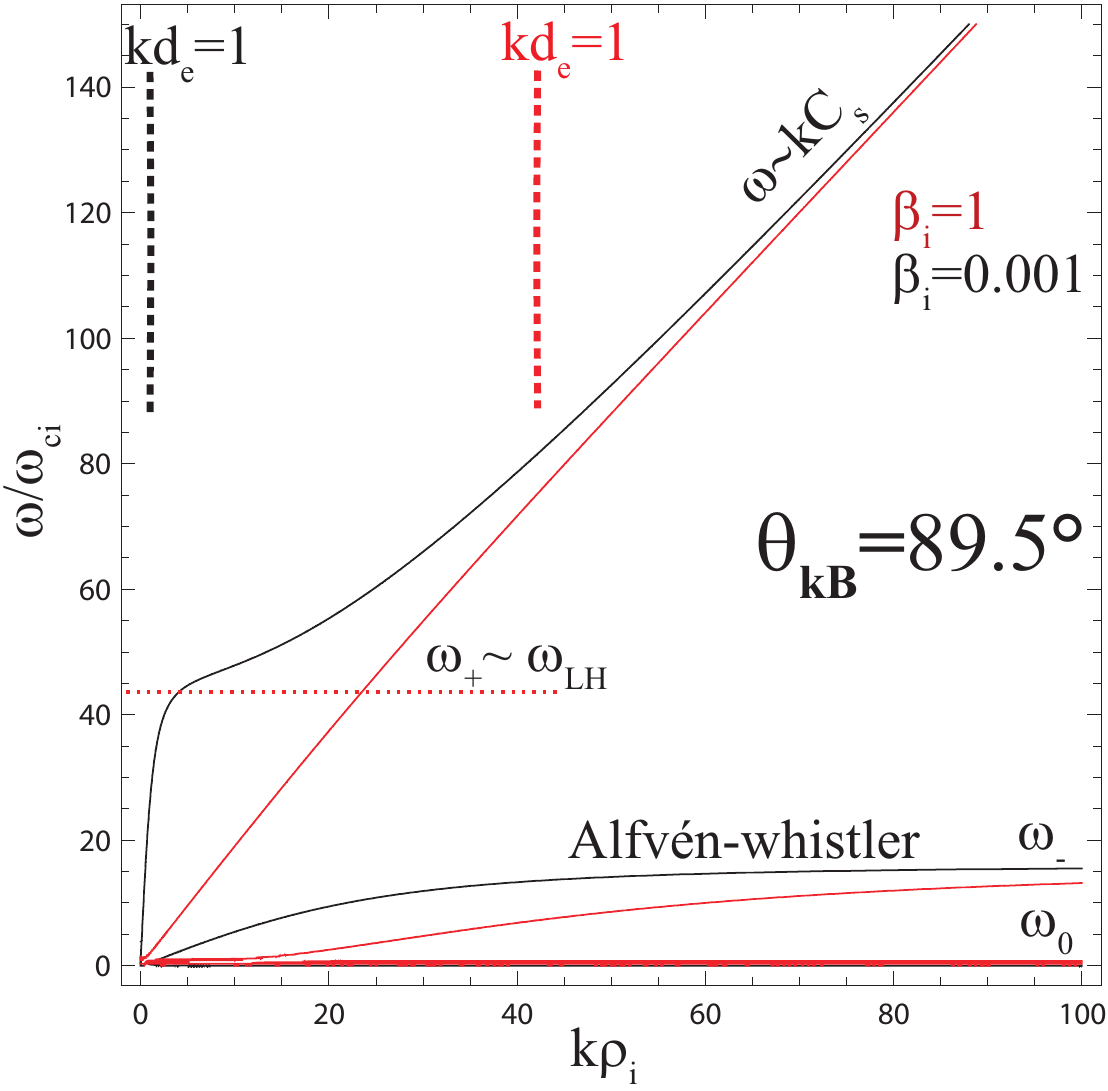}
\caption{Dispersion relations of the reduced-two-fluid model in the hot (red) and cold (black) plasmas with $T_i=T_e$, $\gamma_i=5/3$ and $\gamma_e=1$. The vertical dashed lines correspond to the electron inertial length $kd_e=1$ for the two values of $\beta_i$. The insert in the top panel is a zoom on the very small scales of the plot to show the asymptotes $\omega_+$ and $\omega_-$. \label{bif_apj}}
\end{figure}

\begin{figure}
\includegraphics[height=5cm,width=8cm]{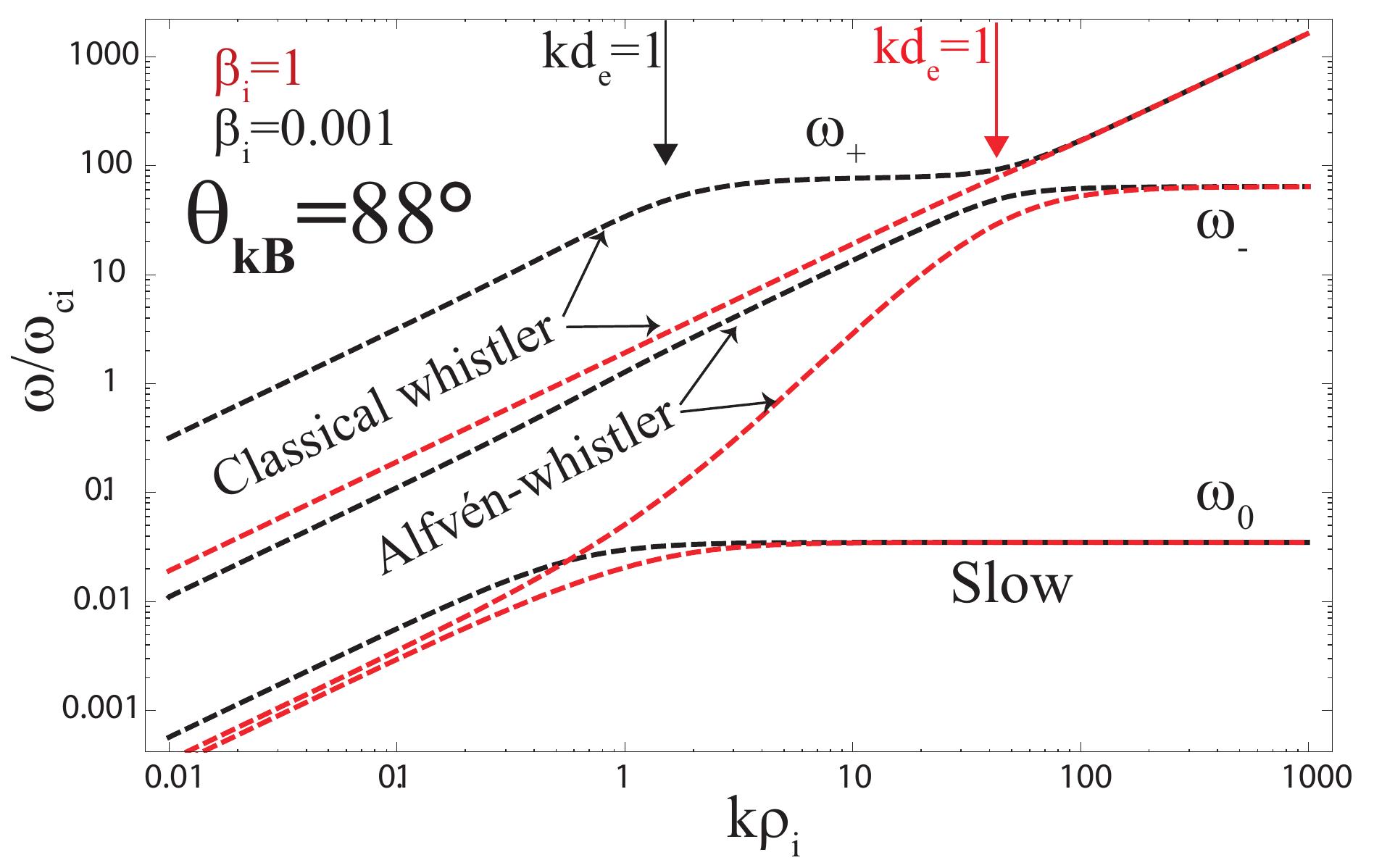}
\caption{Dispersion relations on a log-log scale of the reduced-two-fluid model in the hot (red) and cold (black) plasmas showing the connection between high and low frequency modes (the same description as in Fig.~\ref{bif_apj} applies). \label{bif_log_apj}}
\end{figure}

To derive the linear solutions we linearize equations~(\ref{eq1}-\ref{eq5}) by assuming that $n_s=n_0+\delta n_s, P_s=P_0+\delta P_s, {\bf B}={\bf B}_0+\delta{\bf B}$. If, furthermore, we assume that the perturbations vary as $e^{-j(\omega t-{\bf k\cdot r})}$, one can obtain (after some calculations) the following set of equations
\begin{eqnarray}
\label{eq6}
\delta{\bf E} =B_0 \tens{M}_i \cdot \delta {\bf v}_i \\
\label{eq7}
\delta{\bf E} =B_0 \tens{M}_e \cdot \delta {\bf v}_e \\
\label{eq8}
\delta{\bf v}_i-\delta{\bf v}_e=\frac{1}{B_0}\tens{M} \cdot \delta{\bf E}
\end{eqnarray}
where 
\begin{eqnarray}
\label{eq9}
\tens{ M}_i&=&\bigg[-j\frac{\omega}{\omega_{ci}}\tens{I}+j\gamma_{i}\frac{k^{2}{{V_{th}}_{i}}^{2}}{\omega\omega_{ci}}{\bf e_{k}}{\bf e_{k}}+\tens{ G}\bigg] \\
\label{eq10}
\tens{M}_e&=&\bigg[j\frac{\omega}{\omega_{ce}}\tens{I}-j\gamma_{e}\frac{k^{2}{{V_{th}}_{e}}^{2}}{\omega\omega_{ce}}{\bf e_{k}}{\bf e_{k}}+\tens{G}\bigg] \\
\label{eq11}
\tens{M}&=&j\frac{k^{2}{{V_{A}}_{i}}^{2}}{\omega\omega_{ci}}\bigg[{\bf e_{k}}{\bf e_{k}}-\tens{I}\bigg]
\end{eqnarray}
where $\tens{I}$ is the identity tensor, $\tens{G}={\bf e_ye_x}-{\bf e_xe_y}$ and ${\bf e_k}$ is the unit wave vector assumed to be in the $XZ$ plane 
\begin{displaymath}
{\bf e_k}={\bf k}/k=\left(\begin{array}{c} \sin\theta_{\bf kB}\\ 0\\ \cos\theta_{\bf kB}
\end{array}\right)
\end{displaymath}
${\bf B_0}=B_0{\bf e_z}$,  ${\omega_c}_s=eB/m_s$, ${V_{th}}_s$ and ${V_A}_s$ are the background magnetic field, the gyropulsation, the thermal and the Alfv\'en speeds of the particle $s$,  respectively. Regrouping equations~(\ref{eq9}-\ref{eq11}) we obtain the final equation
\begin{displaymath}
\Big[{\tens{M}_{i}}^{-1}-{\tens{M}_{e}}^{-1}-\tens{M}\Big] \cdot\delta{\bf E}=0
\end{displaymath}
which admits solutions only if 
\begin{equation}
\label{eq12}
Det\Big[\tens{M}_{e}-\tens{M}_{i}-\tens{M}_{e} \cdot \tens{M} \cdot\tens{M}_{i}\Big]=0
\end{equation}
Equation~\ref{eq12} yields the final dispersion equation of the reduced-two-fluid model
\setlength\arraycolsep{2pt}
\begin{eqnarray}
\label{eqdisp}
%& &\cos^4\theta_{\bf k B}(\gamma_i+\gamma_e){k\rho_i}^6 \nonumber\\
%&-&\cos^2\theta_{\bf k B}\Big[(1+\mu)[1+2\beta(\gamma_i+\gamma_e)] \nonumber\\
%& &{k\rho_i}^4+(1+\mu^2)(\gamma_i+\gamma_e)(k\rho_i)^6\Big]{\frac{\omega}{\omega_{ci}}}^2 \nonumber\\
%&+&\bigg[\beta(1+\mu)^2[(1+\cos^2\theta_{\bf k B}+\beta(\gamma_i+\gamma_e)]{k\rho_i}^2+\nonumber\\
%& & \Big[\cos^2\theta_{\bf k B}(1+\mu^3) +\mu(1+\mu)[1+2\beta(\gamma_i+\gamma_e) \nonumber \\
%& & \Big]{k\rho_i}^4+\mu^2(\gamma_i+\gamma_e){k\rho_i}^6\bigg]{\frac{\omega}{\omega_{ci}}}^4 \nonumber \\
%&-& (1+\mu)[\beta^2(1+\mu^2)+2\beta\mu(1+\mu){k\rho_i}^2+\mu^2{k\rho_i}^4]{\frac{\omega}{\omega_{ci}}}^6
& &c^4\gamma X^6 - c^2\Big[(1+\mu_{ei})(1+2\beta_i \gamma) X^4+(1+\mu_{ei}^2)\gamma X^6\Big]Y^2 \nonumber\\
&+&\bigg[\beta_i(1+\mu_{ei})^2(1+c^2+\beta_i \gamma)X^2+ \Big[c^2(1+\mu_{ei}^3) \nonumber\\
&+&\mu_{ei}(1+\mu_{ei})(1+2\beta_i \gamma) \Big]X^4 +\mu_{ei}^2 \gamma X^6\bigg]Y^4 - (1+\mu_{ei}) \nonumber \\
& & \Big[{\beta_i}^2(1+\mu_{ei}^2)+2\beta_i\mu_{ei}(1+\mu_{ei})X^2+\mu_{ei}^2X^4\Big]Y^6 =0
\end{eqnarray}
where $X=k\rho_i$, $Y=\omega/\omega_{ci}$, $\rho_i=V_{th_{i}}/\omega_{ci}$, $\beta_i={V_{th_{i}}}^2/{V_A}^2$, $c=\cos \theta_{\bf k B}$, $\mu_{ei}=m_e/m_i$ and $\gamma=\gamma_e+\gamma_i$ is the total polytropic index. The solutions of equations~\ref{eqdisp} are plotted in Figs.~\ref{bif_apj}-\ref{bif_log_apj} in low and high $\beta_i$ plasmas for the given angles of propagation and SW parameters.

At low frequencies ($\omega<\omega_{ci}$) we found that the slow magnetosonic mode in both low and high $\beta$ has an asymptotic frequency
\begin{displaymath}
\omega_0\sim \omega_{ci}\cos\theta_{\bf kB}
\end{displaymath}
At high frequencies ($\omega>\omega_{ci}$) both hot and cold plasmas have two modes with different asymptotes. Fig.~\ref{bif_log_apj} shows a similar dispersion curve in log-log scale. One can see that the first mode is connected at low frequency ($\omega<\omega_{ci}$) to the shear Alfv\'en mode (known in MHD) and at $kd_e>>1$ develops an asymptote $\omega_{-}$ given by 
\begin{displaymath}
\omega_{-}\sim \omega_{ce}\cos\theta_{\bf kB}
\end{displaymath}

It is worth recalling here that the asymptote $\omega_{-}$ is usually attributed to the whistler mode, which in this case is connected at low frequency to the Alfv\'en mode and {\it not} to the fast magnetosonic mode. Therefore, in the following we will refer to this mode as the {\it Alfv\'en-whistler} mode for the sake of clarity.  We recall here that the designation of slow and fast modes is based upon the ordering in the phase speeds of the modes. While this ordering remains valid in fluid theories at low and high frequency, it can be totally violated in kinetic theory due to possible interconnections between low and high frequency modes~\citep{krauss94,li01}. Figure~\ref{bif_log_apj} shows also that the second HF mode is connected at low frequency (both in high and low $\beta_i$) to the fast magnetosonic mode. Therefore, we will refer  to this mode as the {\it classical-whistler} mode. In the low $\beta_i$ limit the classical-whistler mode undergoes significant curvature near $\omega_{+}$ at all angles of propagation before converging toward the magnetosonic dispersion $\omega\sim k{C_s}$, where $C_s=\sqrt{T_{ei}\gamma_e +\gamma_i}{V_{th}}_i$ is the sound speed and $T_{ei}=T_e/T_i$. From equation~\ref{eqdisp} one can easily show that in the cold case ($\beta_i=0$) the ``asymptote" $\omega_{+}$ is given by
\begin{displaymath}
\omega_{+}\sim \sqrt{{\omega_{ce}}^2\cos^2\theta_{\bf kB}+\omega_{ci}\omega_{ce}}
\end{displaymath}
which has two interesting limits (for $\omega_{ci}<<\omega_{ce}$)
\begin{itemize}
\item $$\lim_{\theta_{\bf kB}\rightarrow 0} \omega_{+} \sim \lim_{\theta_{\bf kB}\rightarrow 0} \omega_{-}=\omega_{ce}$$
\item $$\lim_{\theta_{\bf kB}\rightarrow \frac{\pi}{2}} \omega_{+}=\sqrt{\omega_{ci}\omega_{ce}}$$
\end{itemize}

The first limit shows that the branch $\omega_{+}$ tends toward the known whistler asymptote $\omega_{+}\sim \omega_{ce}$ in quasi-parallel propagation as shown in Fig.~\ref{bif_apj} (top panel), while the second limit shows that $\omega_{+}$ tends toward the lower-hybrid frequency $\omega_{LH}\sim \sqrt{\omega_{ci}\omega_{ce}}\sim 42\omega_{ci}$ for quasi-perpendicular propagation. In contrast, and as can be seen in Fig.~\ref{bif_apj}, the asymptote $\omega_{-}$ of the Alfv\'en-whistler mode continues decreasing as $\theta_{\bf kB}\rightarrow \pi/2$ in both low and high $\beta_i$ and becomes even smaller than $\omega_{ci}$ at {\it all} scales for the angles of propagation
\begin{displaymath}
 \theta_{\bf kB}>\theta_{crit.}=\cos^{-1}(\mu_{ei})
\end{displaymath}
For the real mass ratio $\mu_{ei}$=1/1836 one obtains $\theta_{crit.}=89.97^\circ$. This implies that at quasi-perpendicular propagation (i.e., $ \theta_{\bf kB}>\theta_{crit.}$) there are {\it two} asymptotes below $\omega_{ci}$: one at extremely low frequency, which is the slow magnetosonic mode discussed above ($\omega_0\sim \omega_{ci}\cos\theta_{\bf kB}$) and the other one is the Alfv\'en-whistler mode with the asymptote $\omega_{-}\sim\omega_{ce}\cos\theta_{\bf kB}$. This is the first important conclusion of this part. It is worth recalling that these two modes were found to be linear solutions of the Hall-MHD equations in the incompressible limit~\citep{sahraoui07, galtier06}, and become degenerate at large (MHD) scales. They were referred to respectively as the Alfv\'en and the whistler modes. In such incompressible fluid theories, the classical-whistler mode is indeed ruled out by the incompressibility assumption (i.e., $\beta \rightarrow \infty$). All turbulence theories built in the framework of those incompressible fluid models are thus based on the slow and Alfv\'en-whistler modes discussed here~\citep{galtier06, galtier08}.  The second important finding from Fig.~\ref{bif_apj} is the major change that the classical-whistler mode undergoes  in high $\beta_i$: when approaching quasi-perpendicular propagation the curvature near $\omega_{+}$ disappears and the dispersion curve tends toward the magnetosonic mode $\omega\sim k{C_s}$. As we will show in the next section, when kinetic effects are considered, the classical-whistler mode does not simply extend above $\omega_{ci}$, but rather splits there into different ion Bernstein modes. Becasue the slow magnetosonic mode is strongly damped in high $\beta$ plasma by kinetic effects, the only relevant mode to carry the energy cascade down to the electron scale is the new Alfv\'en-whistler mode.

%--------------------------------------------------------------------------------------------
% New section 
%--------------------------------------------------------------------------------------------
\section{Linear solutions of the Vlasov-Maxwell equations for high oblique propagation}
\begin{figure}
\includegraphics[height=9cm,width=8.5cm]{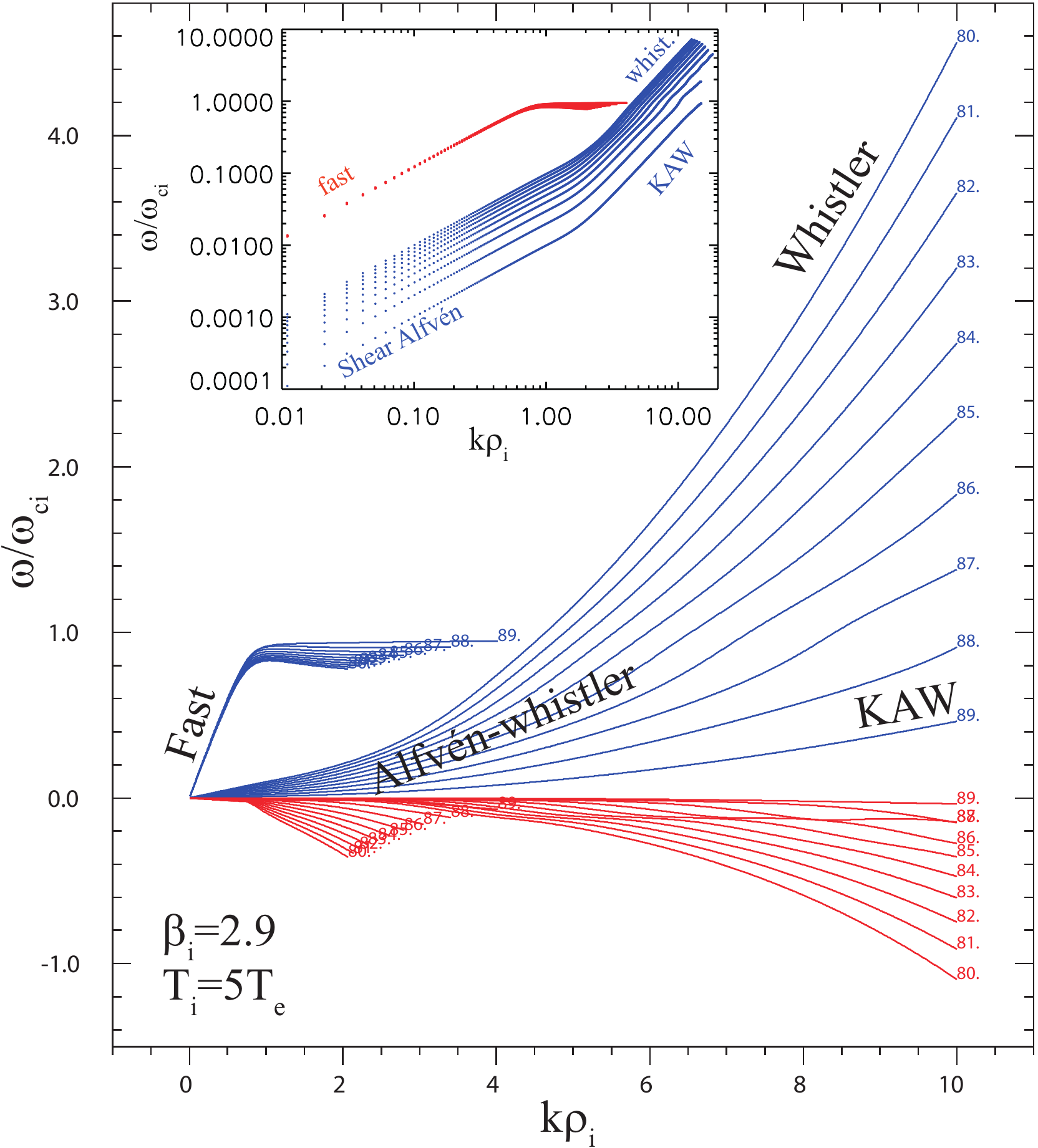}
\caption{Linear solutions of the Maxwell-Vlasov equations: dispersion relations (blue) and damping rates (red) for the angles of propagation $80^\circ\leq\theta_{\bf kB}\leq 89^\circ$.  The insert is a log-log plot of the same dispersion relations to show the connection between low and high frequency modes. \label{whamp1}}
\end{figure}

\begin{figure}
\includegraphics[height=5.5cm,width=8cm]{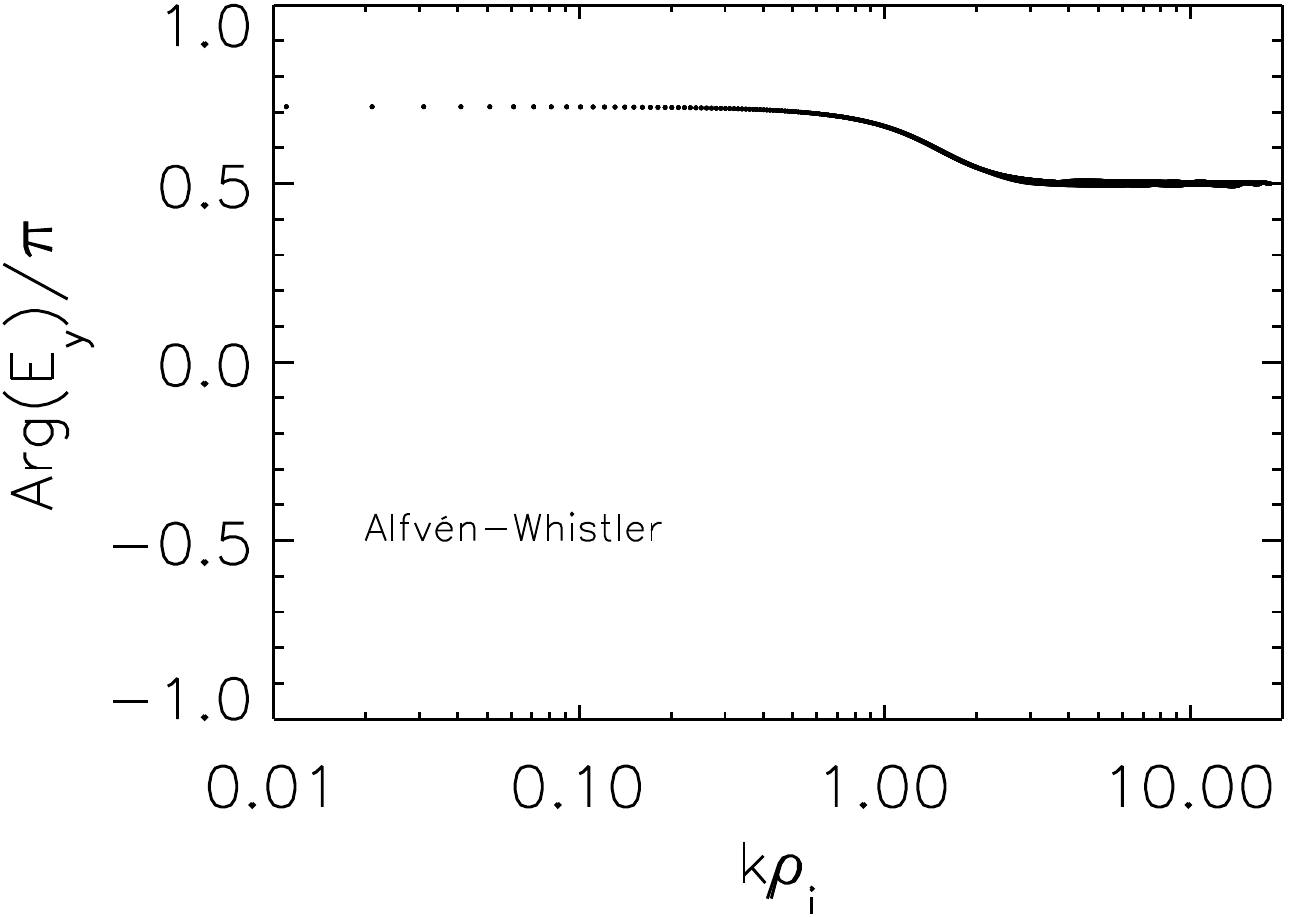}
\includegraphics[height=5.5cm,width=8cm]{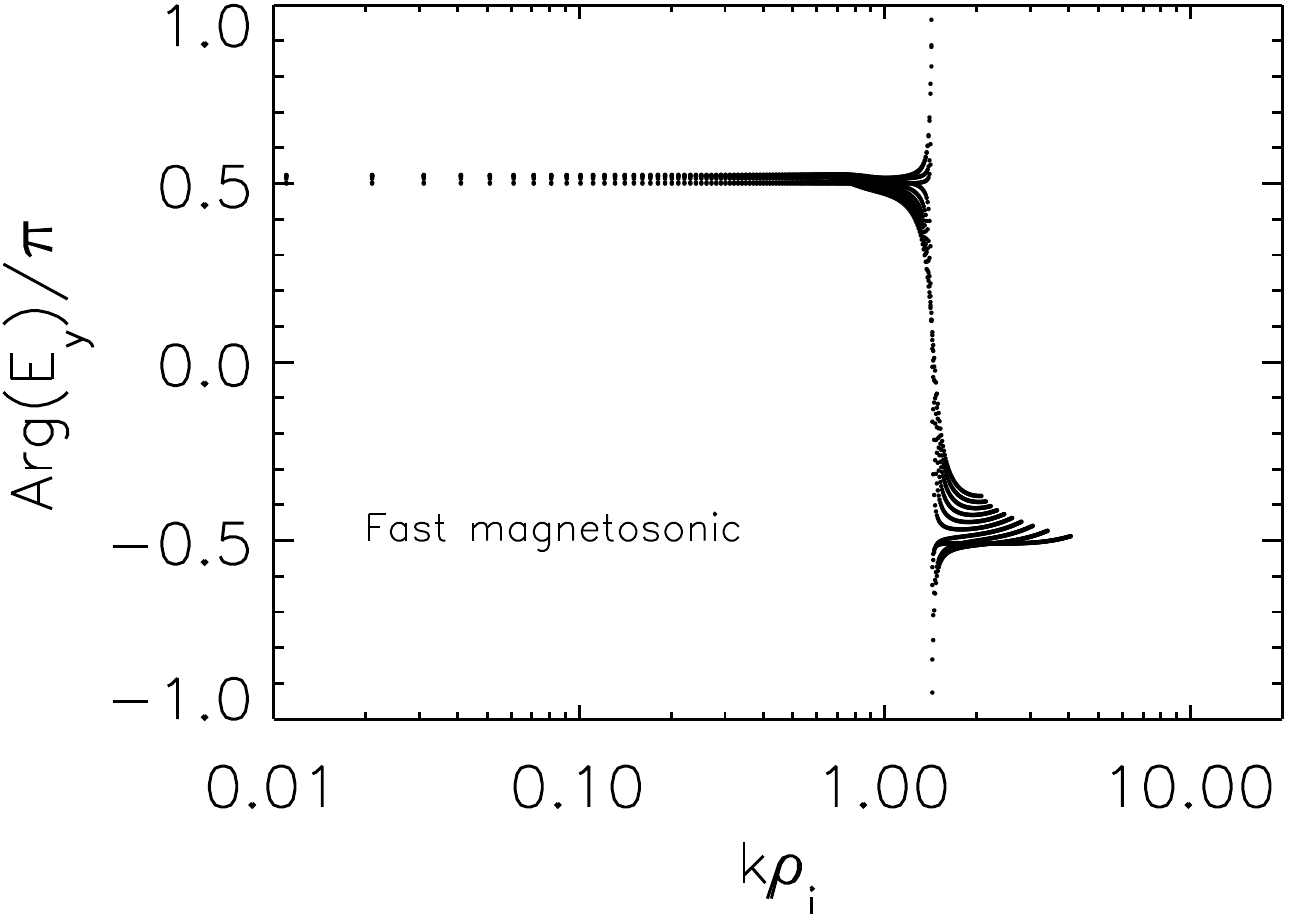}
\caption{Phase of the electric field component $E_y$ of the Alfv\'en-whistler and the fast modes for the angles of propagation $80^\circ\leq\theta_{\bf kB}\leq89^\circ$: it is positive (negative) for right (left) polarized waves. \label{polar}}
\end{figure}

In this section we solve numerically the linear Maxwell-Vlasov equations using the WHAMP code~\citep{ronmark82}, and compare to the previous linear solutions of the reduced hot two-fluid theory. We assume Maxwellian distributions functions of electrons and ions and use realistic SW parameters reported in~\cite{sahraoui09}. Part of the obtained dispersion relations and damping rates are given in Fig.~\ref{whamp1} (the slow magnetosonic mode was found to be heavily damped and thus not plotted here, nor is that mode studied in the rest of this paper). 
The plot shows that the fast magnetosonic modes at high oblique angles undergo resonances at the ion cyclotron frequency, which contrasts with the more known fast mode in quasi-parallel propagation or low $\beta_i$.This result is confirmed by the analysis of the wave polarization. In Fig.~\ref{polar} we plot the phase of the electric field component $E_y$ whose sign gives the sense of polarization:  $Arg(E_y)>0$ (resp. $ <0)$ for right (resp. left) hand polarized waves. We see a clear transition from right to left hand polarization of the fast mode near $k\rho_i\sim 1$ (corresponding to $\omega\sim \omega_{ci}$).  

\begin{figure}
\includegraphics[height=5.5cm,width=8.5cm]{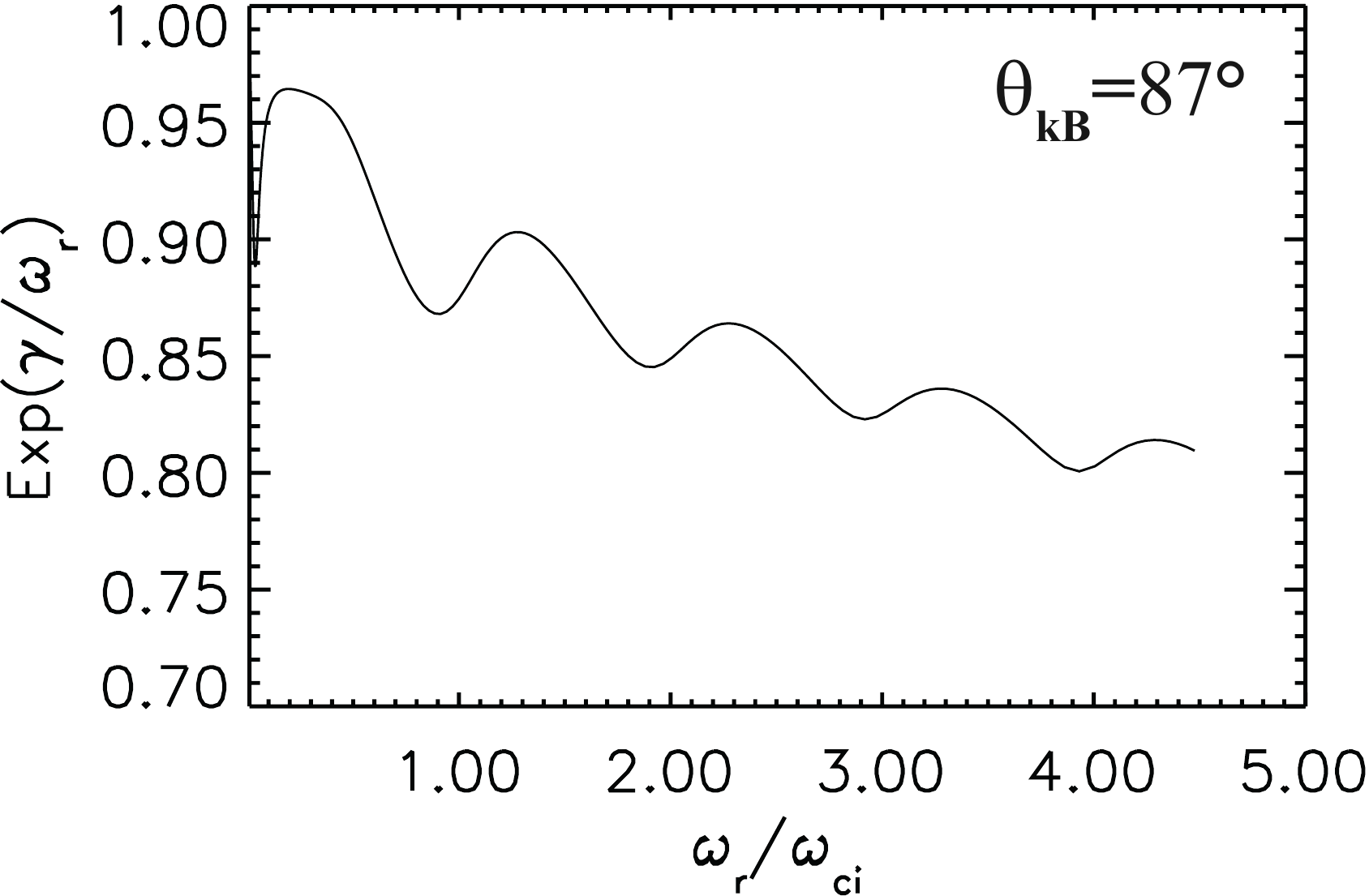}
\caption{Exponential of the normalized damping rate of the Alfv\'en-whistler at $\theta_{\bf kB}=87^\circ$ showing enhacement of the damping near the harmonics of ions. \label{damping_kaw87}}
\end{figure}

In Fig.~\ref{whamp1} we find also the kinetic counterpart of the Alfv\'en-whistler mode derived from the hot-two-fluid theory. The Alfv\'en-whistler mode extends the shear Alfv\'en mode to small scales as can be seen in the insert of Fig.~\ref{whamp1} (this result was also found in Fig. 1 of~\cite{krauss94} although not discussed in detail in that paper). The mode becomes dispersive at scales $k\rho_i\gtrsim 1$ and develops frequencies larger (resp. smaller) than $\omega_{ci}$ for $\theta_{\bf kB}<88^\circ$ (resp. $\geq 88^\circ$) up to the scale $k\rho_i\leq10$. We refer to the branches $\omega<\omega_{ci}$ and $\omega \gtrsim \omega_{ci}$ respectively as the KAW and the whistler modes. Note that the limit $\omega=\omega_{ci}$ is reached at different spatial scales depending on the value of the angle $\theta_{\bf kB}$. Figure~\ref{polar} shows that in contrast to the fast magnetosonic mode, the Alfv\'en-whistler mode has a right hand polarization at all scales and does not undergo significant change near $k\rho_i \sim 1$. 
Nevertheless, the Alfv\'en-whistler modes show features related to wave-particle resonances near the harmonics of ions (because they are not strictly circularly polarized). This can be seen in Fig.~\ref{damping_kaw87} which shows the enhancement of the damping of the Alfv\'en-whistler modes near the harmonics $\omega_N=N\omega_{ci}-k_\parallel V_{th_i}$ (a similar observation can be made on Fig.~\ref{whamp4} at other angles of propagation). 

\begin{figure}
\includegraphics[height=9cm,width=8.5cm]{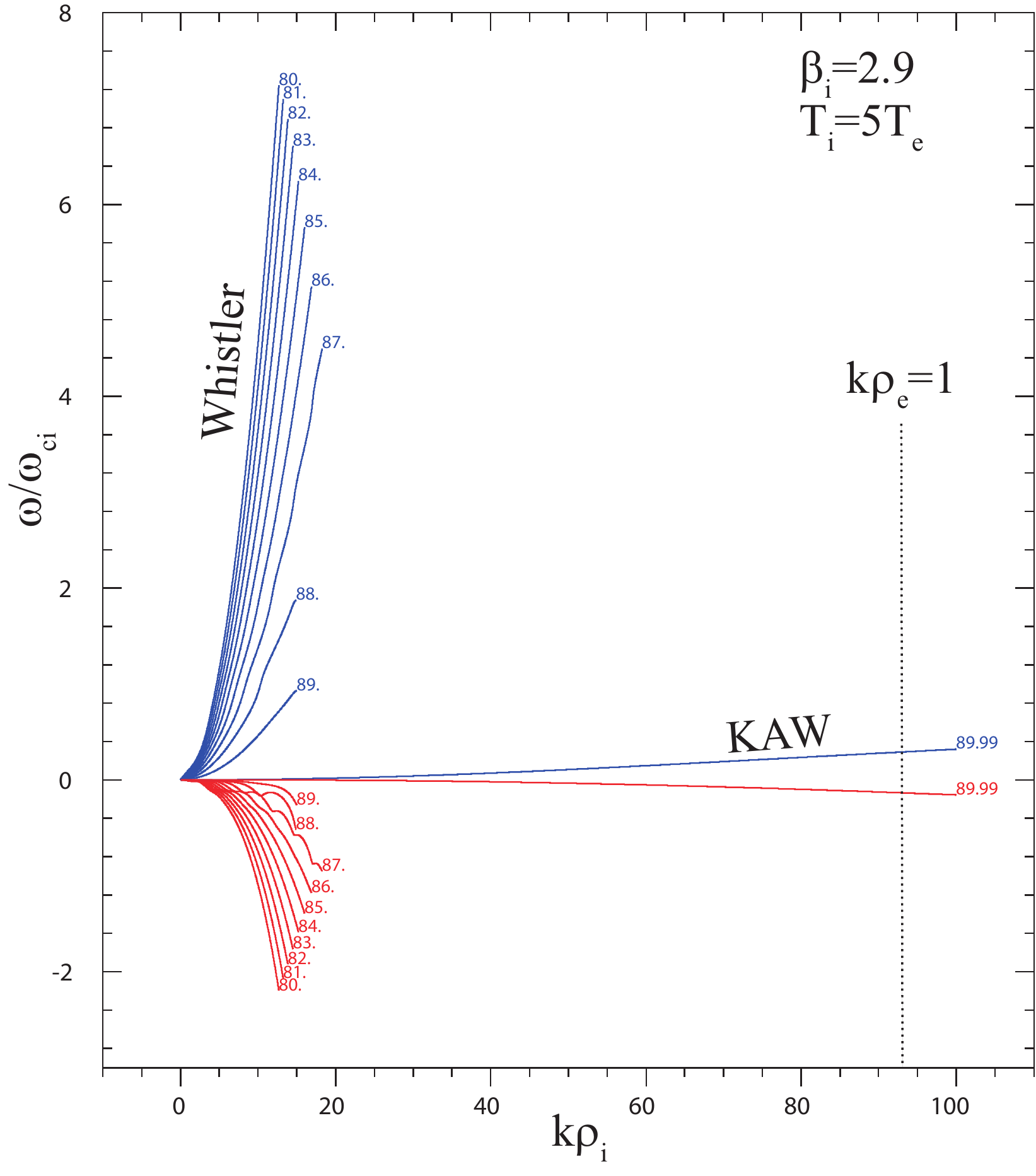}
\caption{Dispersion relations of the Alfv\'en-whistler modes (blue) and their damping rates (red) for the angles of propagation $80^\circ\leq\theta_{\bf kB}\leq89.99^\circ$. \label{whamp2}}
\end{figure}

\begin{figure}
\includegraphics[height=5.5cm,width=8cm]{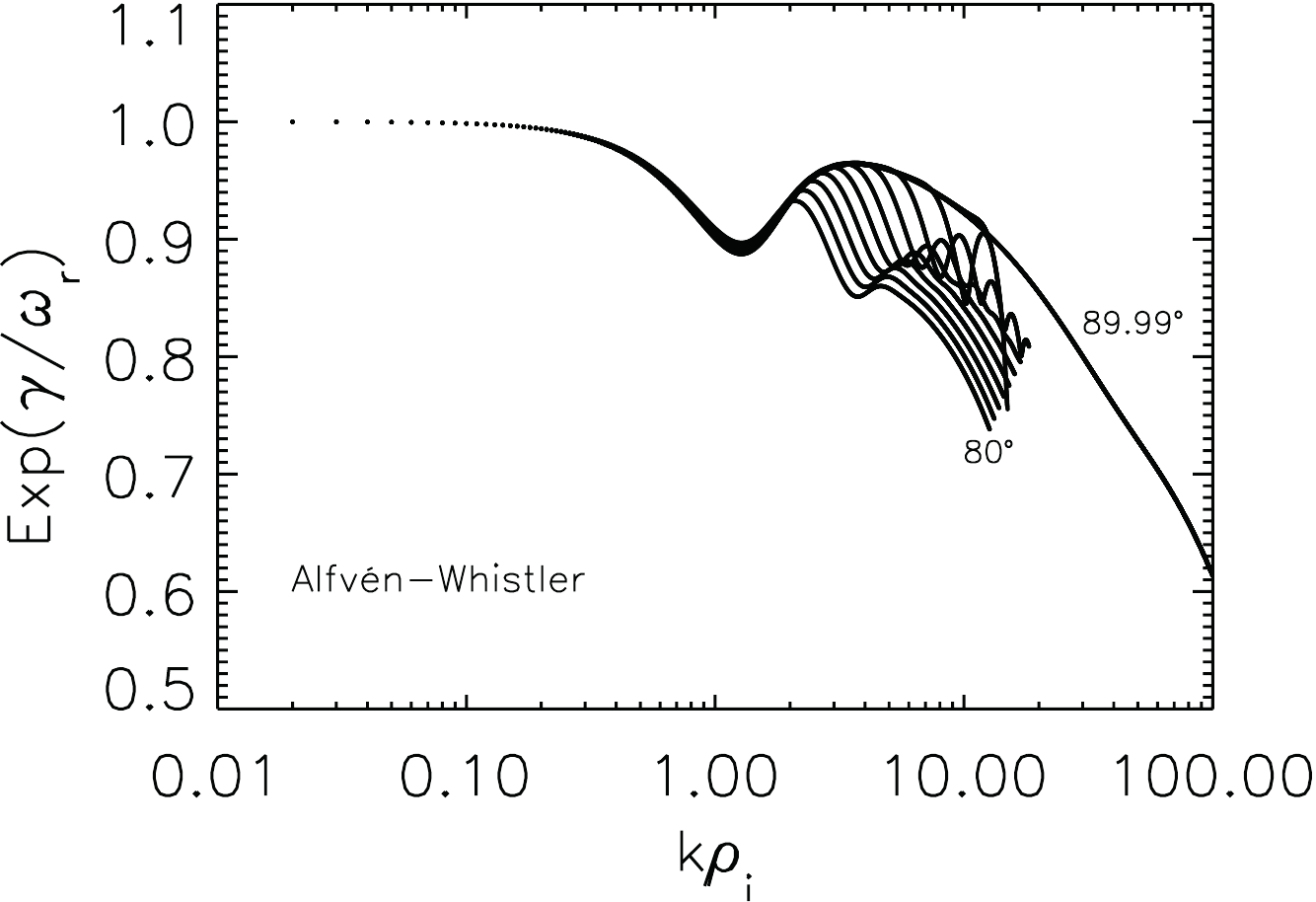}
\includegraphics[height=5.5cm,width=8cm]{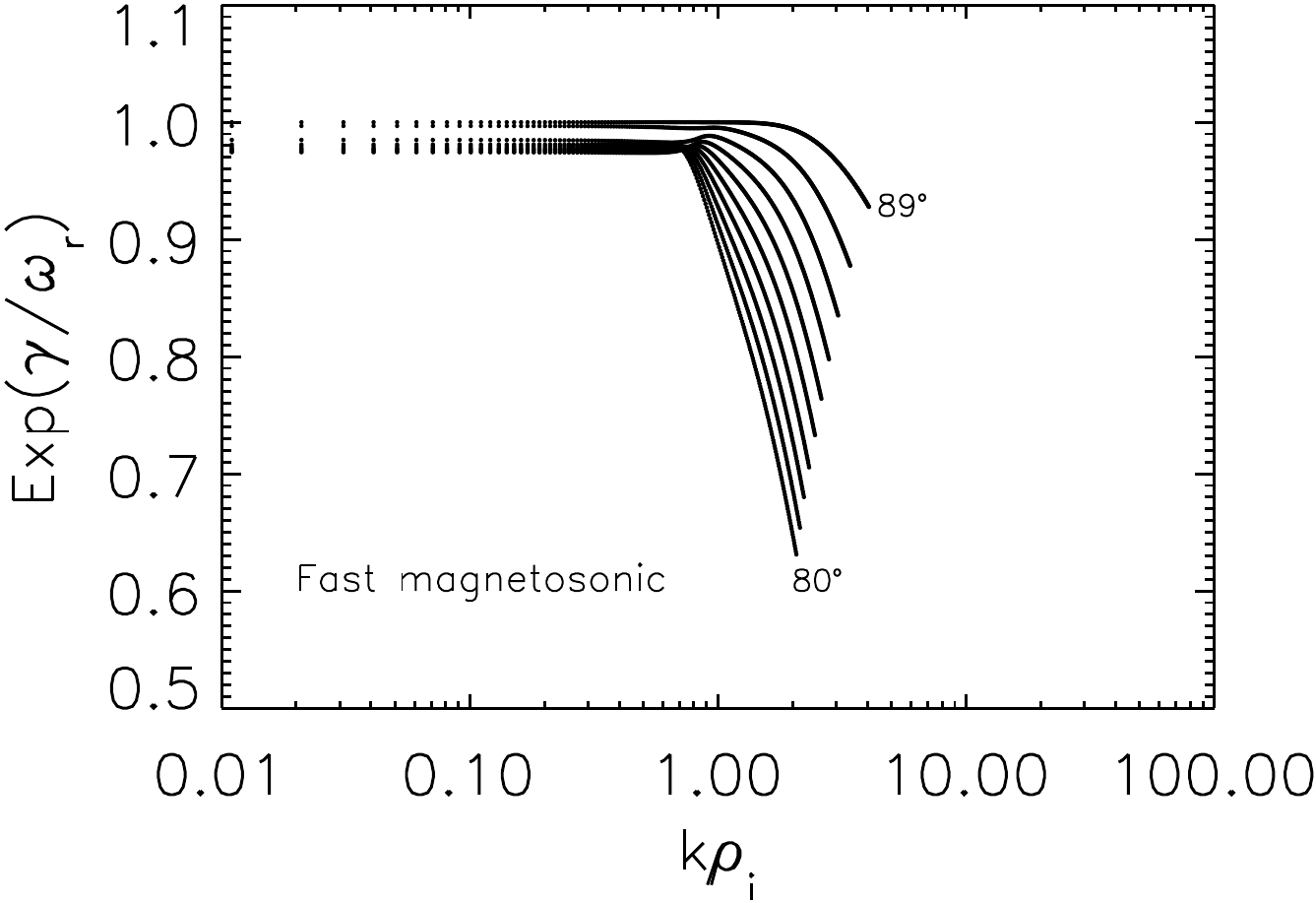}
\includegraphics[height=5.5cm,width=8cm]{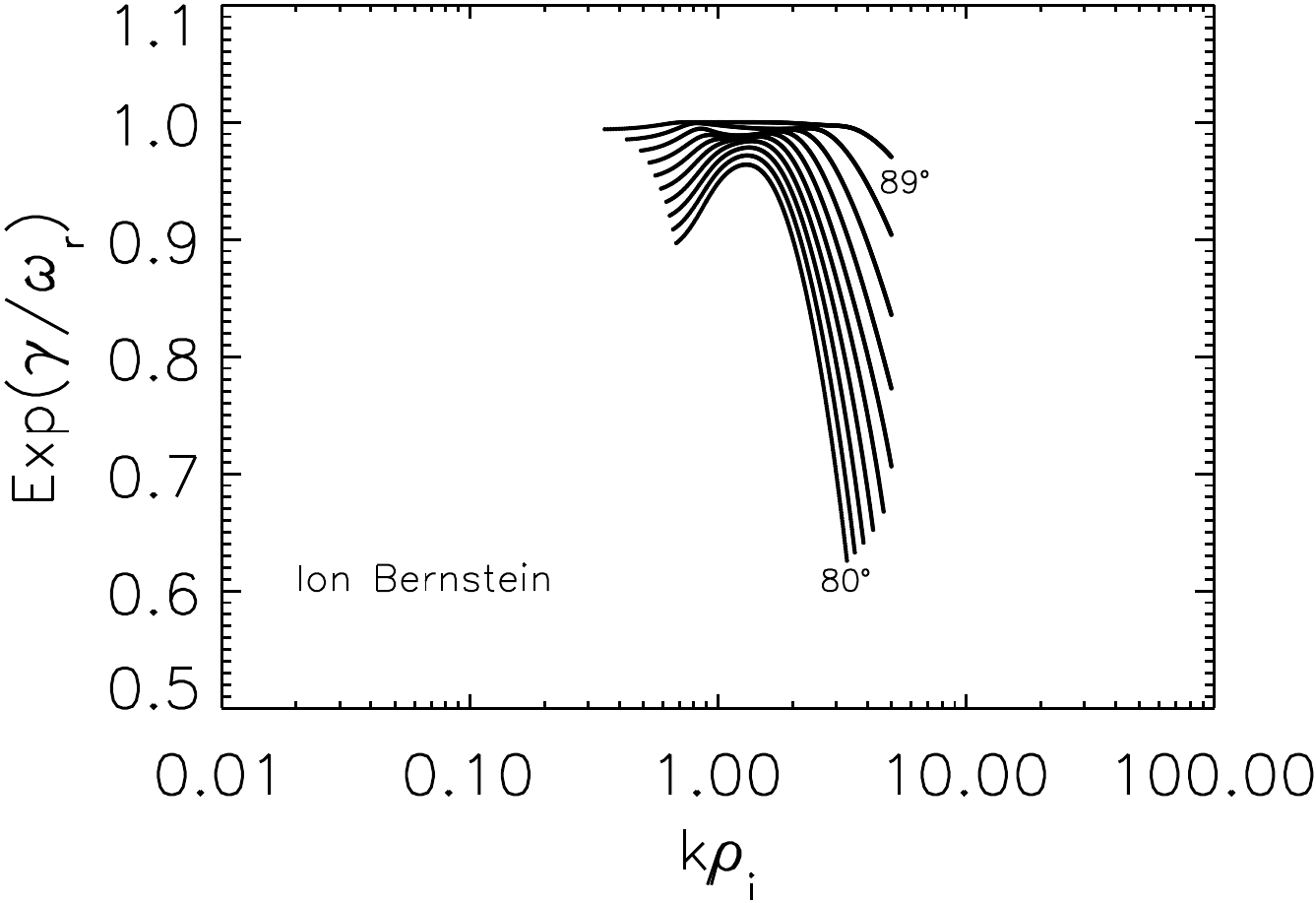}
\caption{Exponential of the damping rates of the Alfv\'en-whistler, fast, and Bernstein modes, normalized to the frequency of each wave for angles of propagation $80^\circ\leq\theta_{\bf kB}\leq 89^\circ$. \label{damping}}
\end{figure}

Fig.~\ref{whamp2} shows the Alfv\'en-whistler solutions extended to high frequencies and small scales. We observe that the damping of the Alfv\'en-whistler mode becomes more important when departing from $\theta_{\bf kB}\sim 90^\circ$ toward less oblique angles. For $\theta_{\bf kB}=89.99^\circ$ the solution extends down to the electron gyroscale $\rho_e$ where the damping rate remains small\footnote{This solution has been used in~\cite{sahraoui09} to interpret SW observations.}, $\gamma/\omega_r\sim 0.4$. For less oblique angles the Alfv\'en-whistler mode develops frequencies higher than $\omega_{ci}$ but they are subject to stronger damping. 
This can be seen clearly in Fig.~\ref{damping}, which shows the damping rate in one period of each wave mode. Figure~\ref{damping} (top panel) shows that the Alfv\'en-whistler modes are abruptly damped at $k\rho_i\sim 1$. At smaller scales the most oblique modes are the least damped. The linear damping rate may thus play a role of a ``filter that lets pass'' only very oblique modes at small scales. This may explain the high oblique modes frequently observed in the SW \citep{sahraoui10a,narita11,sahraoui11}.

\begin{figure}
\includegraphics[height=9cm,width=8.5cm]{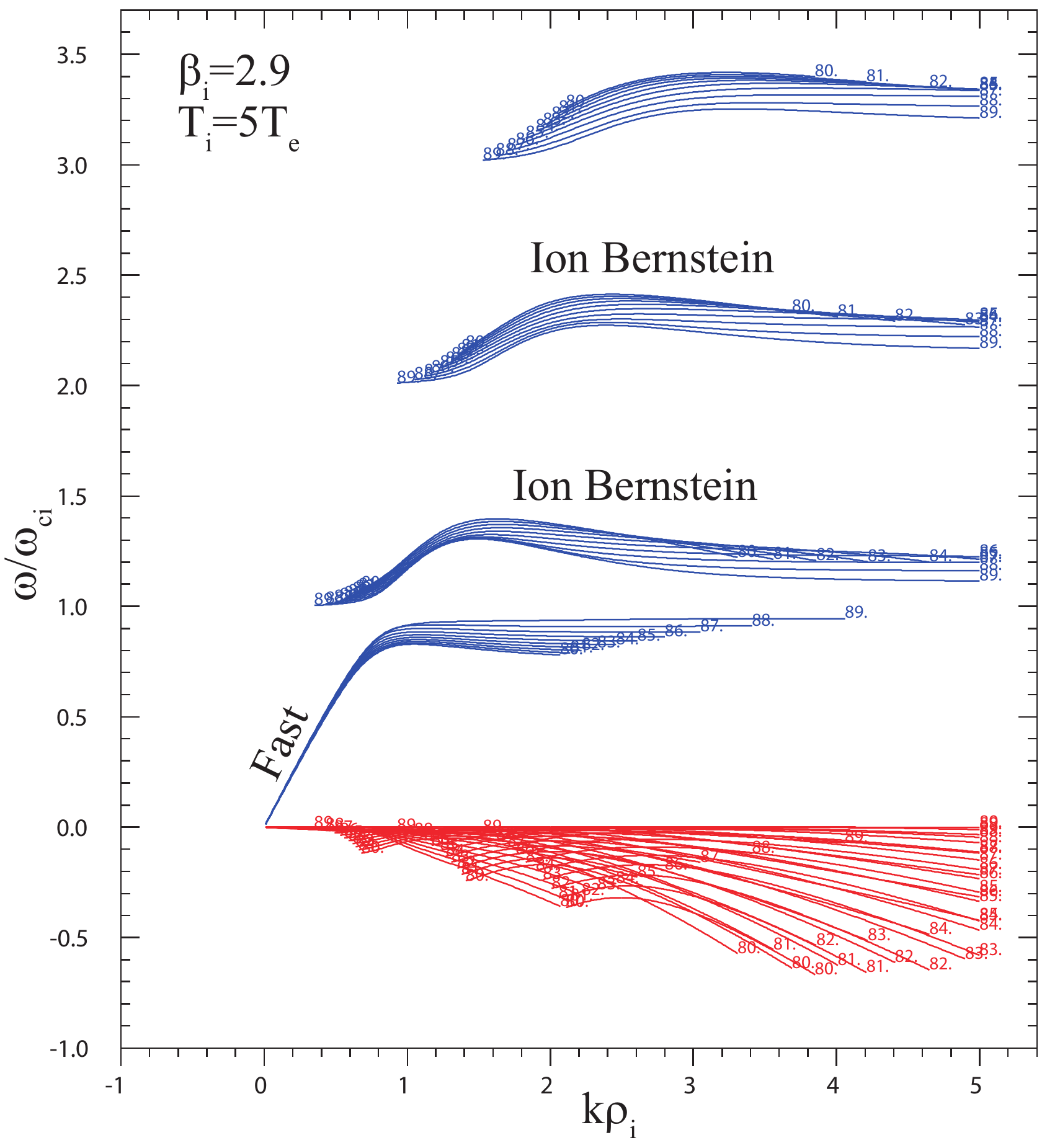}
\caption{Dispersion relations of the fast and the ion Bernstein modes (blue) and their damping rates (red) for the angles of propagation $80^\circ\leq\theta_{\bf kB}\leq 89^\circ$. \label{whamp3}}
\end{figure}

\begin{figure}
\includegraphics[height=7.5cm,width=7.5cm]{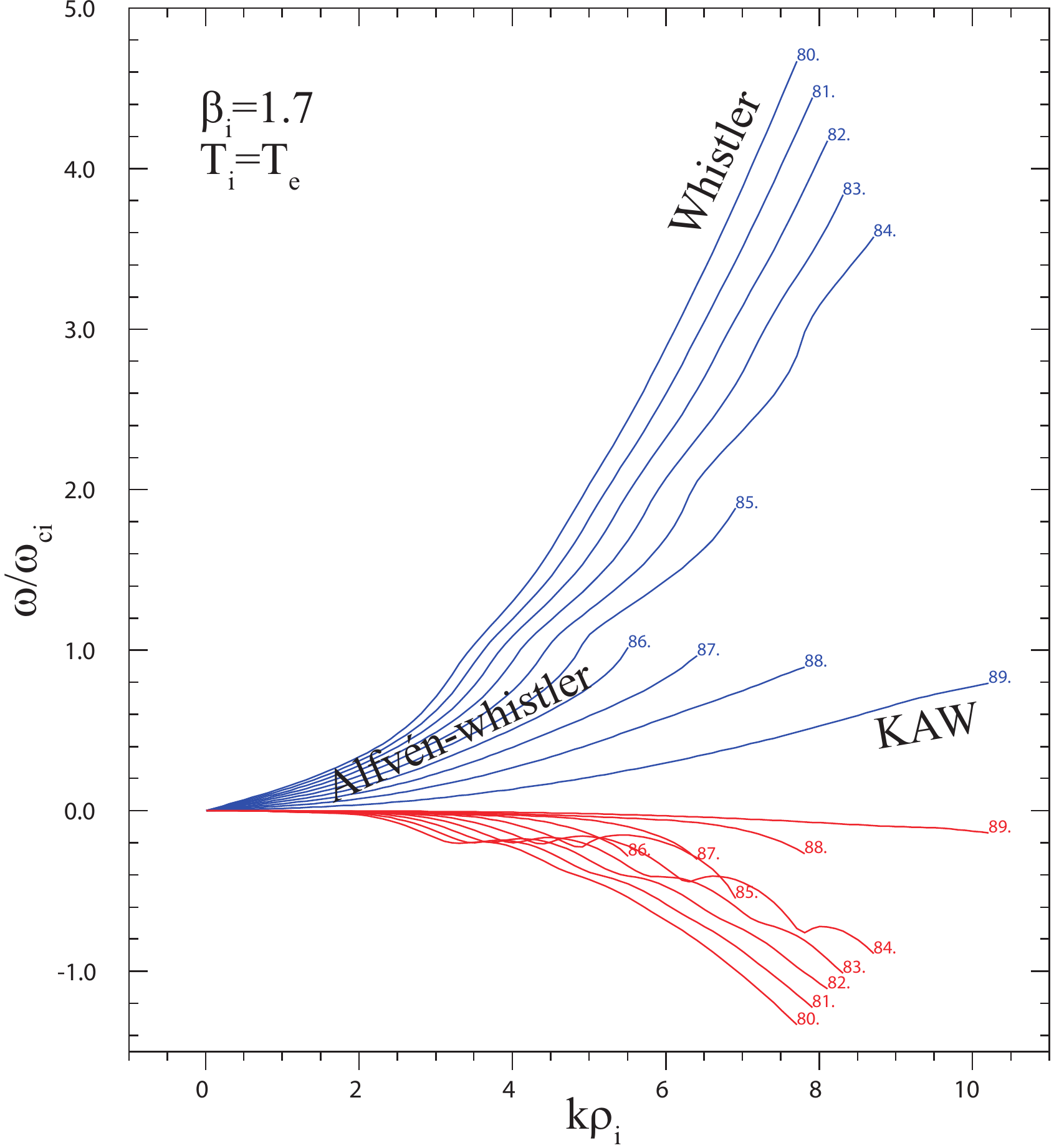}
\includegraphics[height=7.5cm,width=7.5cm]{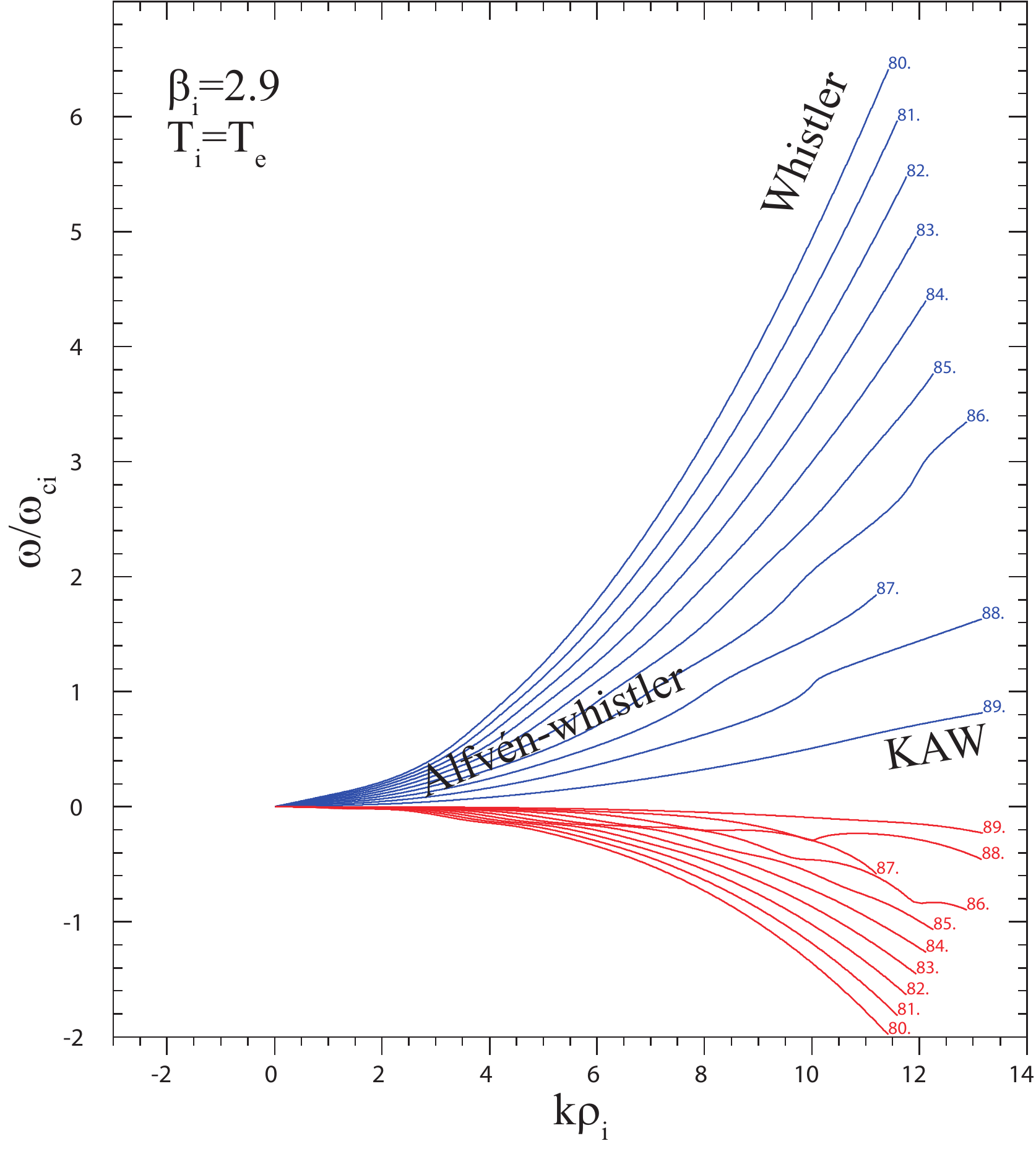}
\includegraphics[height=7.5cm,width=7.5cm]{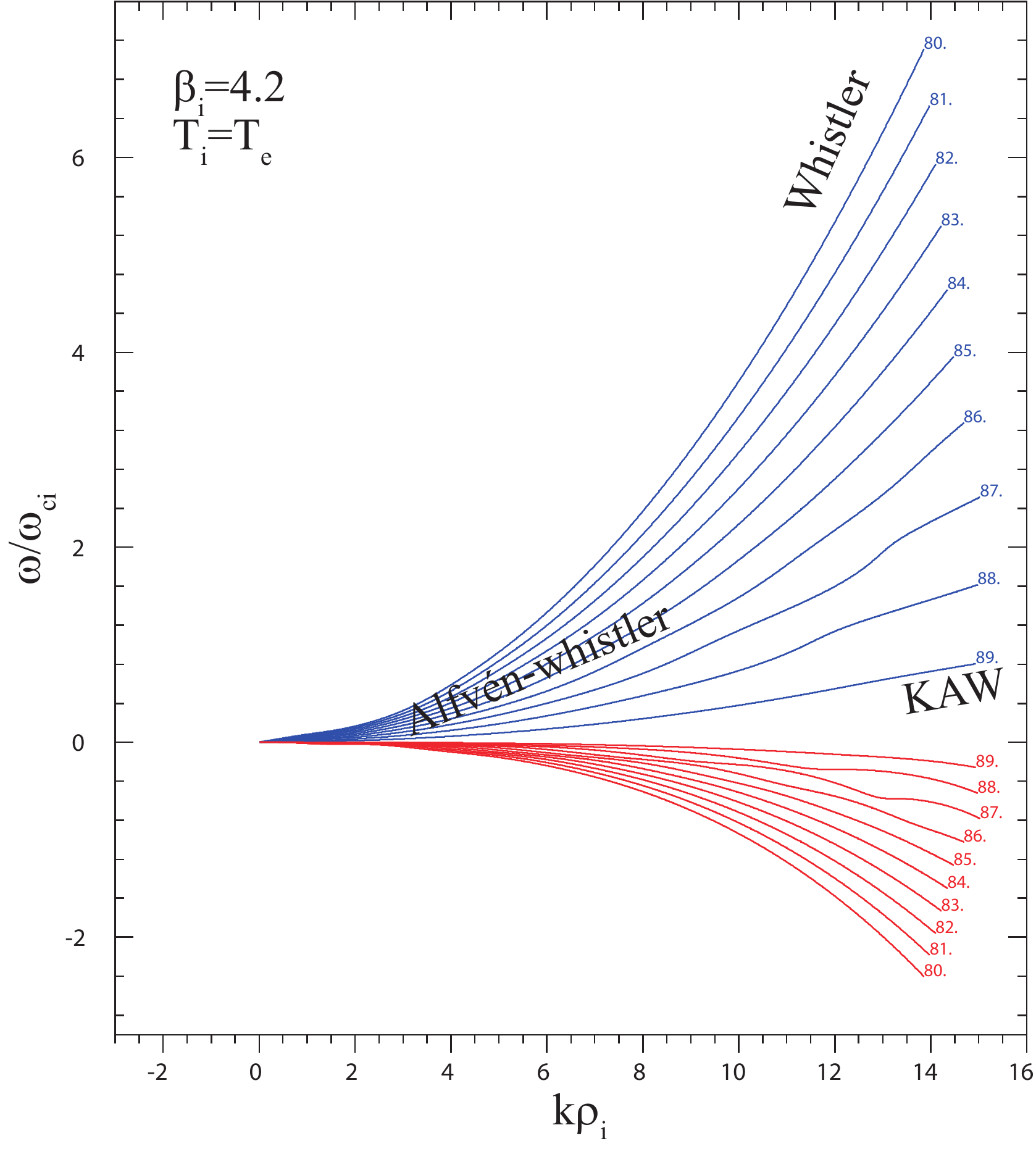}
\caption{Dispersion relations of the Alfv\'en-whistler modes (blue) and their damping rates (red) for the angles of propagation $80^\circ\leq\theta_{\bf kB}\leq89^\circ$ and for different values of $\beta_i$ (all have $T_i=T_e)$. \label{whamp4}}
\end{figure}

 A similar extension of the fast mode solutions to frequencies higher than $\omega_{ci}$ is shown in Fig.~\ref{whamp3}. We observe frequency gaps in the dispersion curves caused by the ion Bernstein modes, known to develop in hot plasmas at quasi-perprendicular angles of propagation (e.g., \cite{li01}). The more oblique Bernstein modes are the least damped as shown in Fig.~\ref{damping} (lower panel). This result clearly invalidates the magnetosonic dispersion $ \omega\sim kC_s$ of the classical-whistler mode at $\omega >\omega_{ci}$ found in the previous section from the hot two-fluid theory.

Figure~\ref{damping} shows clearly that, while the Bernstein and the fast modes are less damped than the Alfv\'en-whistler modes near $k\rho_i\sim1$, both the Bernstein and the fast magnetosonic modes become heavily damped at $k\rho_i\gtrsim 3$. The Alfv\'en-whistler modes appear to be the less damped ones at scales $k\rho_i\geq3$. Furthermore, among these modes the most oblique one is the least damped, which is the KAW mode at $\theta_{\bf kB}=89.99^\circ$ in Fig. \ref{damping}. 

In Fig.~\ref{whamp4} we tested the validity of these results for different high $\beta_i$ values and with $T_i=T_e$.  We see clearly that the conclusions above remain valid and that the ratio $T_i/T_e$ does not modify much the results but slight changes of the dispersion curves near the harmonics of ions. This can be seen for instance by comparing Fig.~\ref{whamp4} (middle panel) to Fig.~\ref{whamp1} which was plotted for the same $\beta_i=2.9$ but with $T_i=5T_e$. It is important however to note that when $\beta_i$ is becoming smaller (i.e. decreasing toward 1) the damping of Alfv\'en-whistler modes by cyclotron resonances become important for some angles of propagation, as can be seen in Fig.~\ref{whamp4} (top panel). We tested other values of $\beta_i$ (not shown here) and found that the conclusions above remain valid. Moreover, we found that higher values of $\beta_i$ allow extending the Alfv\'en-whistler modes to even smaller scales (for $\beta_i=\beta_e=25$, the highest tested value, the modes were found to extend up to $k\rho_i\sim 30$). On the contrary, for $\beta_i\sim 1$ we found that none of the Alfv\'en-whistler modes (when $80^\circ\leq\theta<90^\circ$) can propagate at $\omega>\omega_{ci}$ because they are strongly damped at $\omega\sim \omega_{ci}$. 

\begin{figure}
\includegraphics[height=5.5cm,width=7cm]{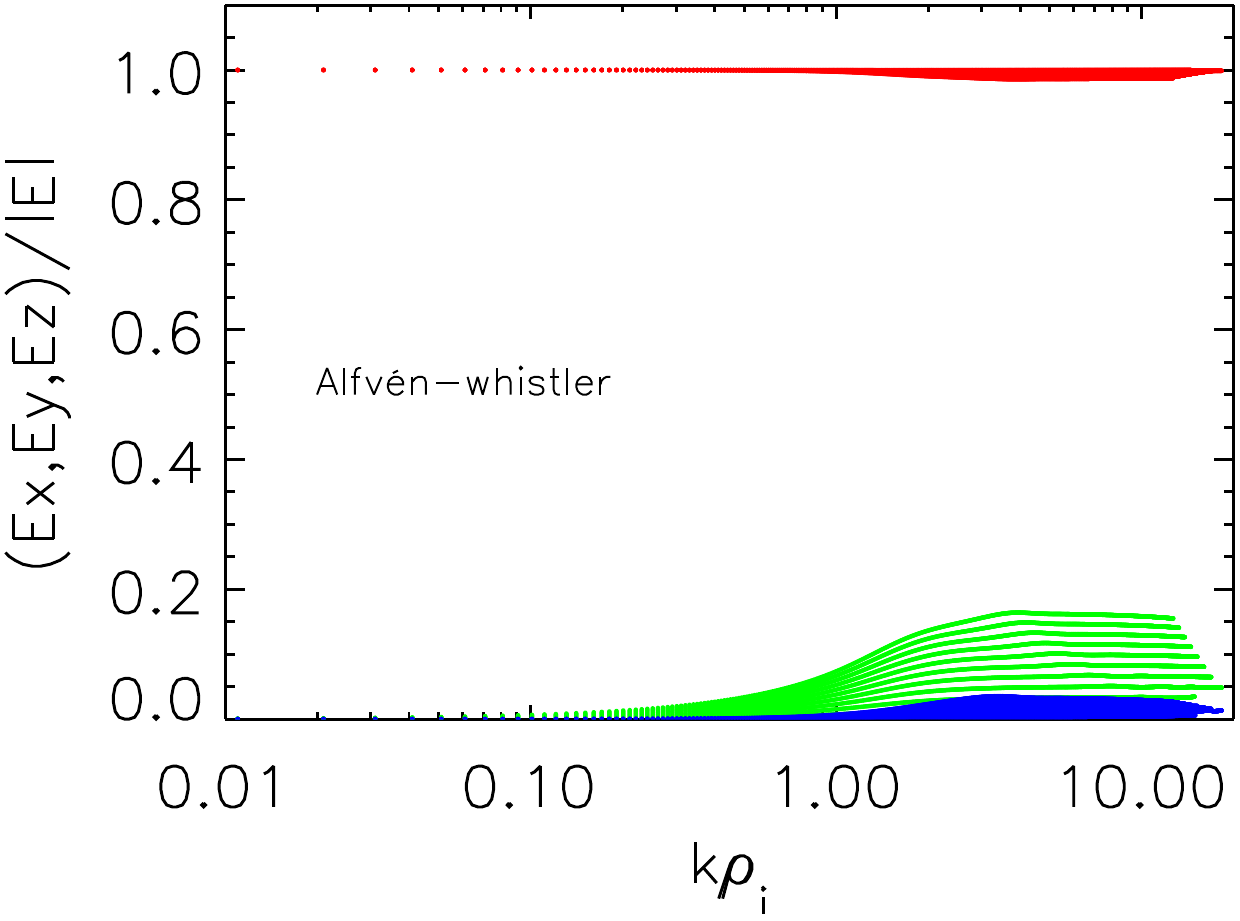}
\includegraphics[height=5.5cm,width=7cm]{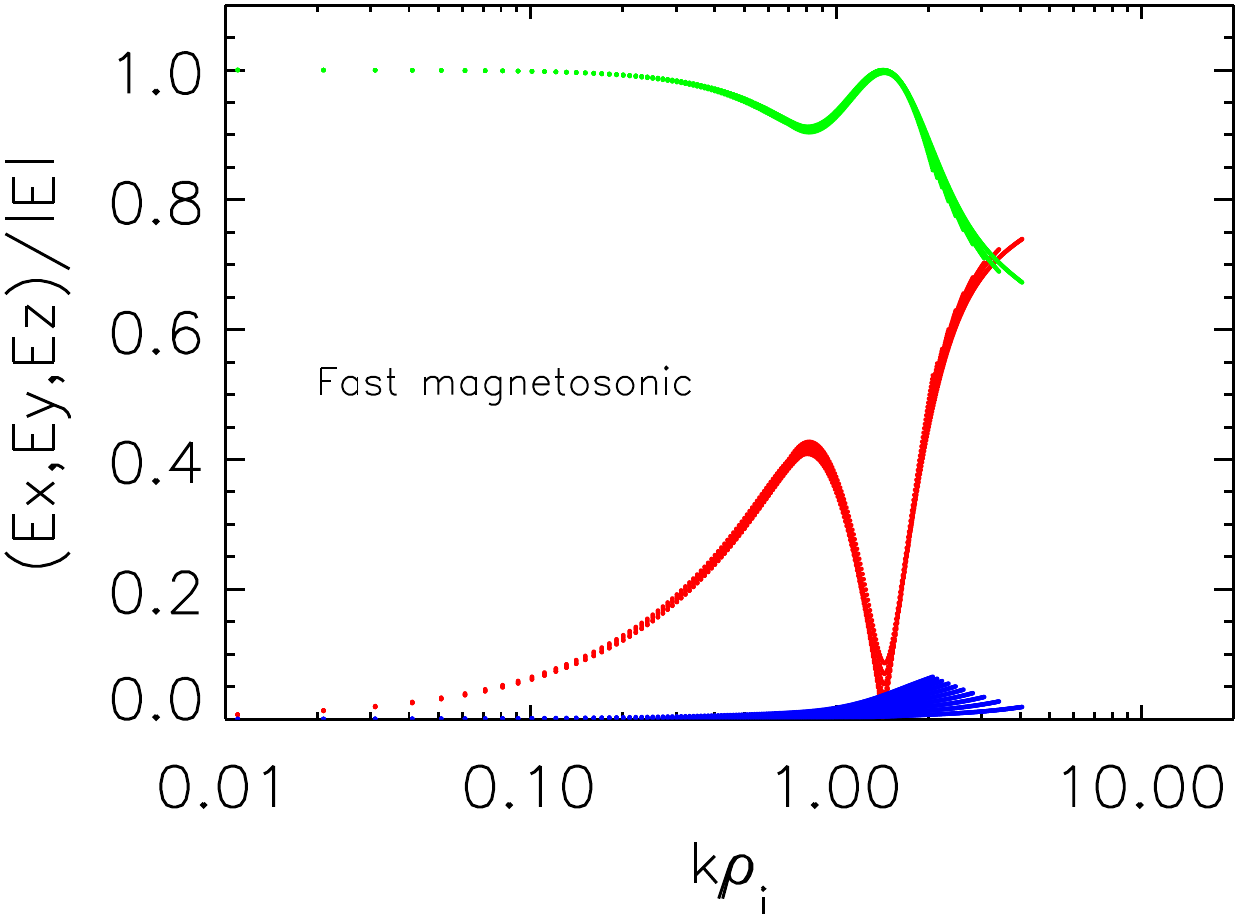}
\includegraphics[height=5.5cm,width=7cm]{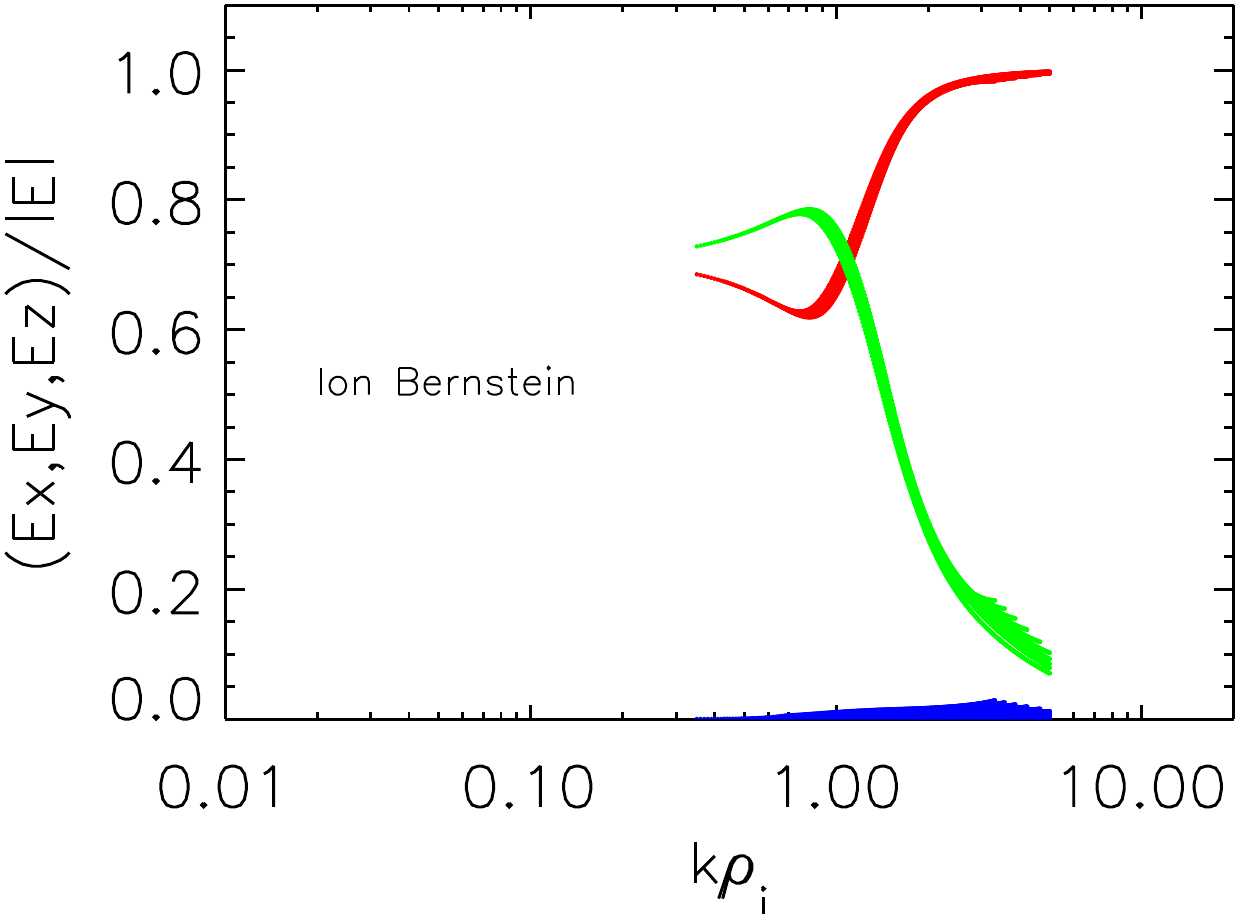}
\caption{Electric field components $E_x$ (red), $E_y$ (green) and $E_z$ (blue) of the Alfv\'en-whistler, fast magnetosonic and the ion Bernstein modes normalized to their total electric field, computed for the angles of propagation $80^\circ\leq\theta_{\bf kB}\leq89^\circ$. The $E_x$ component (red) is essentially the electrostatic part of the electric field in each mode, while $E_z$ (blue) is the parallel component (to ${\bf B}_0$) of the electric field. \label{elect}}
\end{figure}

Now one can ask the question: which of these plasma modes, fast, Bernstein, KAW or whistler, is likely to carry the energy cascade of turbulence down to  the dissipation scales in the limited range of SW parameters studied here?
While all these modes might contribute to the energy cascade in the SW, based on linear damping rates and assuming that (quasi-)linear theory is applicable to small scale SW fluctuations, we can conclude that the KAW branch ($\omega<\omega_{ci}$) is more likely to be observed in the data than the whistler branch ($\omega>\omega_{ci}$). However, damping rates alone cannot rule out the presence of fast or ion Bernstein modes in the transition range $0.5\leq k\rho_i\leq 3$ where they are found to be less damped than the Alfv\'en-whistler modes. Why, then, rule out the fast and the Bernstein modes in this range of scales? It has been often argued~\citep[e.g., ][]{howes09} that the Bernstein modes are highly electrostatic and therefore cannot account for the magnetic spectra observed in the dispersion range in the SW. This is not totally correct as can be seen in Fig.~\ref{elect}, which shows the three components of the electric field of each mode. With ${\bf k}$ chosen in the $XZ$ plane and considering the high oblique angles studied here ($80^\circ\leq\theta_{\bf kB}\leq89^\circ$) the $E_X$ component (in red in the plot) represents essentially the electrostatic field. It is instructive to observe that the electromagnetic part of the Bernstein mode dominates over the electrostatic part at scales $0.2\leq k\rho_i\leq1$ and that the electrostatic component dominates only at smaller scales. More interestingly, we observe that the Alfv\'en-whistler modes are essentially electrostatic at {\it all} scales. This, in fact, is not surprising: in the MHD limit the field perturbations of the shear Alfv\'en mode $\delta {\bf E}$ and $\delta {\bf B}$ and the mean field ${\bf B}_0$ are orthogonal to each other. At quasi-perpendicular angles the electric field $\delta {\bf E}$ is therefore necessarily quasi-longitudinal (i.e., parallel to ${\bf k}$).  

Figure~\ref{elect} shows that only the fast mode has a dominant electromagnetic field up to scales $k\rho_i \gtrsim3$ where it becomes comparable to the electrostatic part. The three modes have very small parallel electric field (given by the $E_z$ component in the plot). Therefore, Fig.~\ref{elect} shows clearly that one cannot rule out any mode only on the basis of the strength of its electrostatic component.  Instead we suggest using the magnetic compressibility~\citep{song94,lacombe95}, in addition to the damping rates discussed above, to rule out the fast and the Bernstein modes in SW observations. 

\begin{figure}
\includegraphics[height=5cm,width=7.5cm]{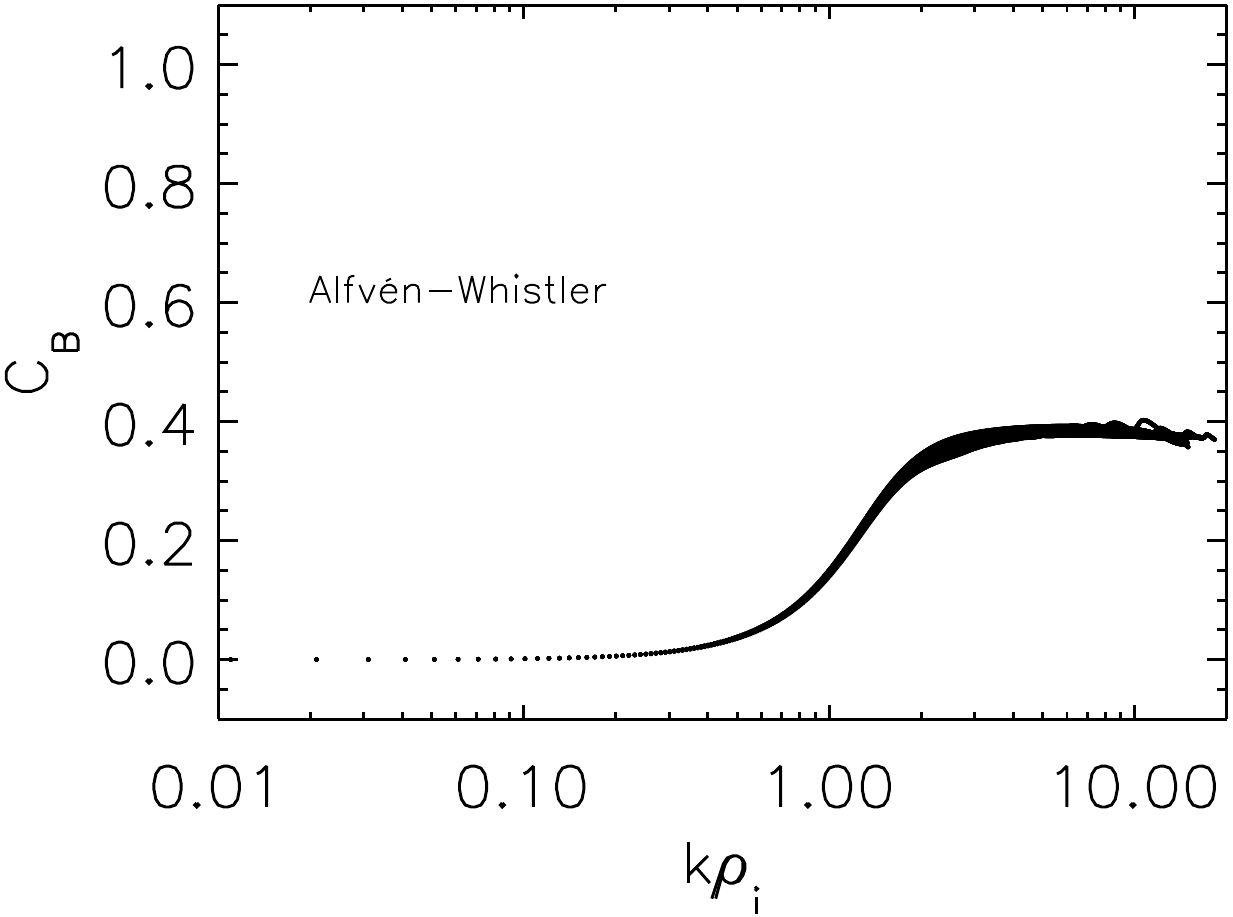}
\includegraphics[height=5cm,width=7.5cm]{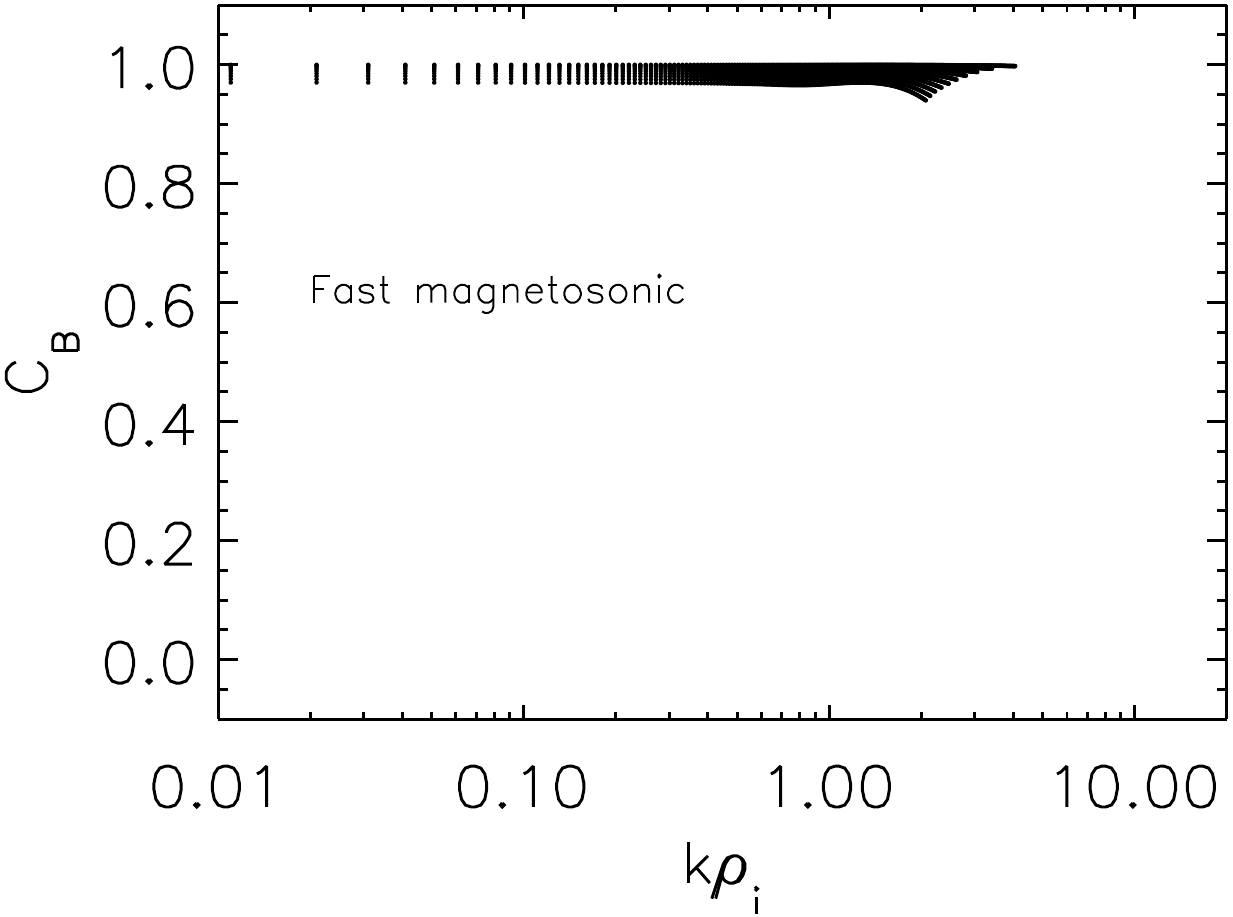}
\includegraphics[height=5cm,width=7.5cm]{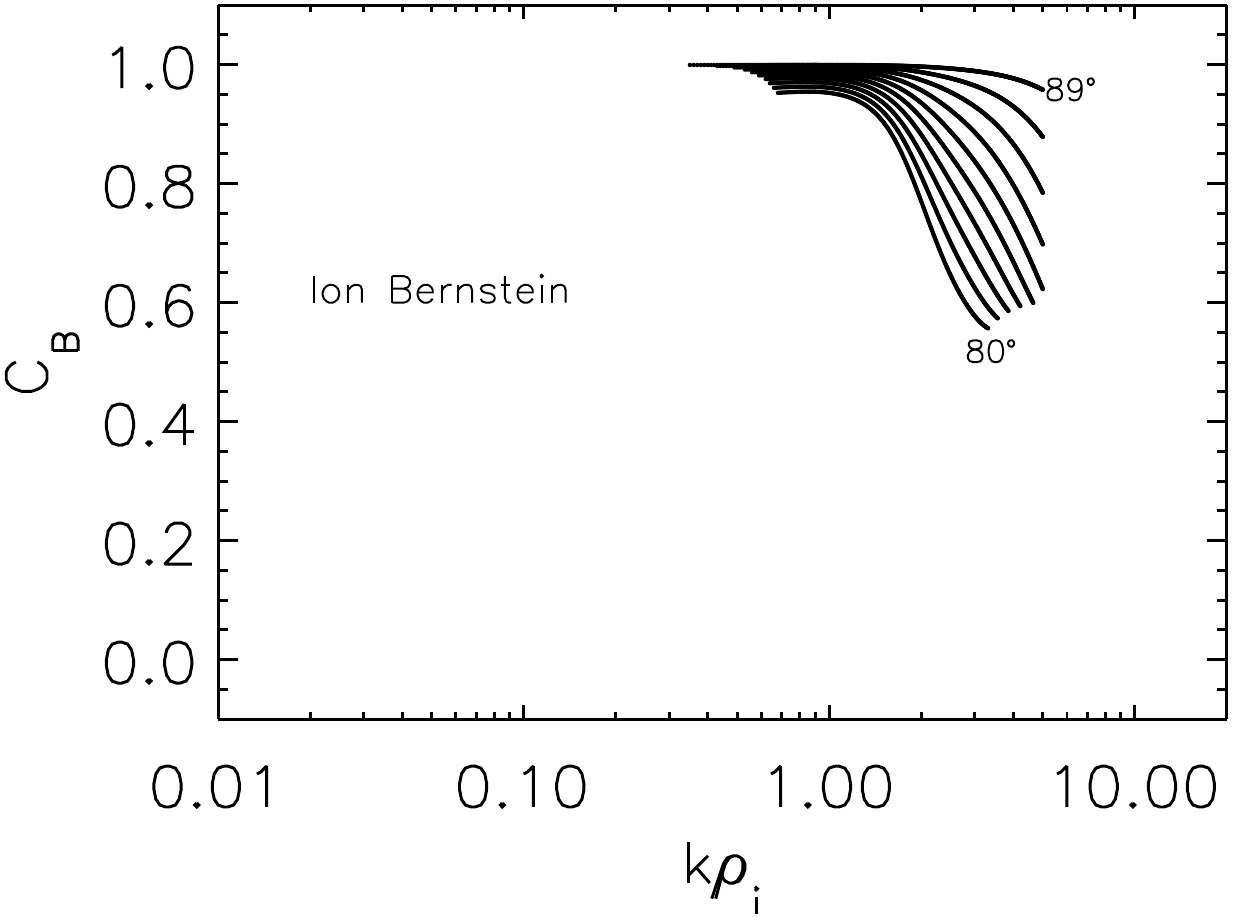}
\caption{Magnetic compressibility of the Alfv\'en-whistler, fast, and Bernstein modes for the angles of propagation $80^\circ\leq\theta_{\bf kB}\leq89^\circ$. \label{compres}}
\end{figure}

We define the magnetic compressibility by $C_B={\delta B_\parallel}^2/{\delta B}^2$ (where ${\delta B}^2$ is the total magnetic power). It is a well established result that SW magnetic compressibility is small in the inertial range, i.e., $C_B \sim 0.1$, then it increases to $C_B \sim 0.4$ in the transition and the dispersive ranges~\citep{sahraoui10a,kiyani11}. In Fig.~\ref{compres} we plot the magnetic compressibilities of the three modes for the same angles of propagation as before. The magnetic compressibility profile of each mode shows clearly that the Alfv\'en-whistler modes fit SW observations better, at least at the scales $k\rho_i\gtrsim 1$, since both the Bernstein and the fast modes show the dominance of parallel power over perpendicular power  ($C_B\sim1$). However, it is important to notice in Fig.~\ref{compres} (top panel) that all the Alfv\'en-whistler modes (at different angles) have the same profile of magnetic compressibility. This implies that, at the oblique angles studied here, one cannot answer the question as to which branch, KAW or the whistler, dominates in the data soley from the measurement of the magnetic compressibility. This contrasts with the conclusions given in \cite{gary09}. As discussed above, the computed damping rates showed that the whistler branch is more damped than the KAW branch at small scales.

%--------------------------------------------------------------------------------------------
% New section 
%--------------------------------------------------------------------------------------------
\section{Discussion}
The previous results show that in high $\beta_i$ plasma and at high oblique angles of propagation, there is one solution that extends from the MHD to the electron scales: it is the Alfv\'en-whistler mode. The whistler mode discussed here develops as a continuation at high frequency of the shear Alfv\'en mode  and {\it  not} of the fast magnetosonic mode. This result rises some exceptions to the general claim made by \cite{howes11b} on the absence of whistler modes in small scale SW turbulence because of the absence of fast magnetosonic modes in the inertial range.  The fast mode strongly resonates with ions near $\omega_{ci}$ and splits into different ion Bernstein modes at higher frequencies. In contrast, the shear Alfv\'en mode extends at scales $k\rho_i\gtrsim1$ to frequencies either larger or smaller than $\omega_{ci}$, depending on the ratio $k_\parallel/ k_\perp$. The {\it same} mode can thus be called whistler ($\omega>\omega_{ci}$) or KAW ($\omega<\omega_{ci}$) depending on this ratio. If $k_\parallel/k_\perp<\mu_{ei}$ then the Alfv\'en mode follows the KAW mode with $\omega<\omega_{ci}$ at all scales (even at $k\rho_e> 1$). For larger values of $k_\parallel/k_\perp$, the Alfv\'en mode develops a whistler branch with $\omega \gtrsim \omega_{ci}$ at a given scale, but the damping rate increases as well. Based on the damping rates one would expect to observe more oblique KAW than oblique whistlers. This result may be considered as another, rather simpler, alternative to explain the strong anisotropies ($k_\perp>>k_\parallel)$ observed in the SW. Indeed, there are several predictions from different MHD and kinetic turbulence models, e.g., critical balance conjecture~\citep{goldreich95}, that predict stronger anisotropy at small scales~\citep{shebalin83}. Here we argue that the {\it linear} kinetic damping of the Alfv\'en-whistler modes as discussed above may as well explain such observed strong anisotropy: the more oblique Alfv\'en-whistler modes are indeed the less damped ones. 

From this result it appears clear that the question as to which branch, KAW or whistler, dominates in the data reduces to the question as to how oblique is the wave vector ${\bf k}$. The answer requires accurate measurement of the angle $\theta_{\bf kB}$, which in turn requires {\it simultaneous} measurement of $k_\parallel$ and $k_\perp$. This is now possible using the {\it k}-filtering technique on Cluster spacecraft data \citep{sahraoui06,sahraoui10a, narita10a}. However, as one can see from Fig.~\ref{whamp1}, a change in the angle of propagation from $\theta_{\bf kB}=80^\circ$ to $\theta_{\bf kB}\sim 90^\circ$ will change the physics from the whistler branch to the KAW branch. Measuring the angle $\theta_{\bf kB}$ with an accuracy better than $10^\circ$ is quite difficult, even with the powerful multi-spacecraft techniques, owing to the various sources of uncertainties in the measurements \citep{sahraoui03a,sahraoui10c}. 

\begin{figure}
\includegraphics[height=5.5cm,width=8.5cm]{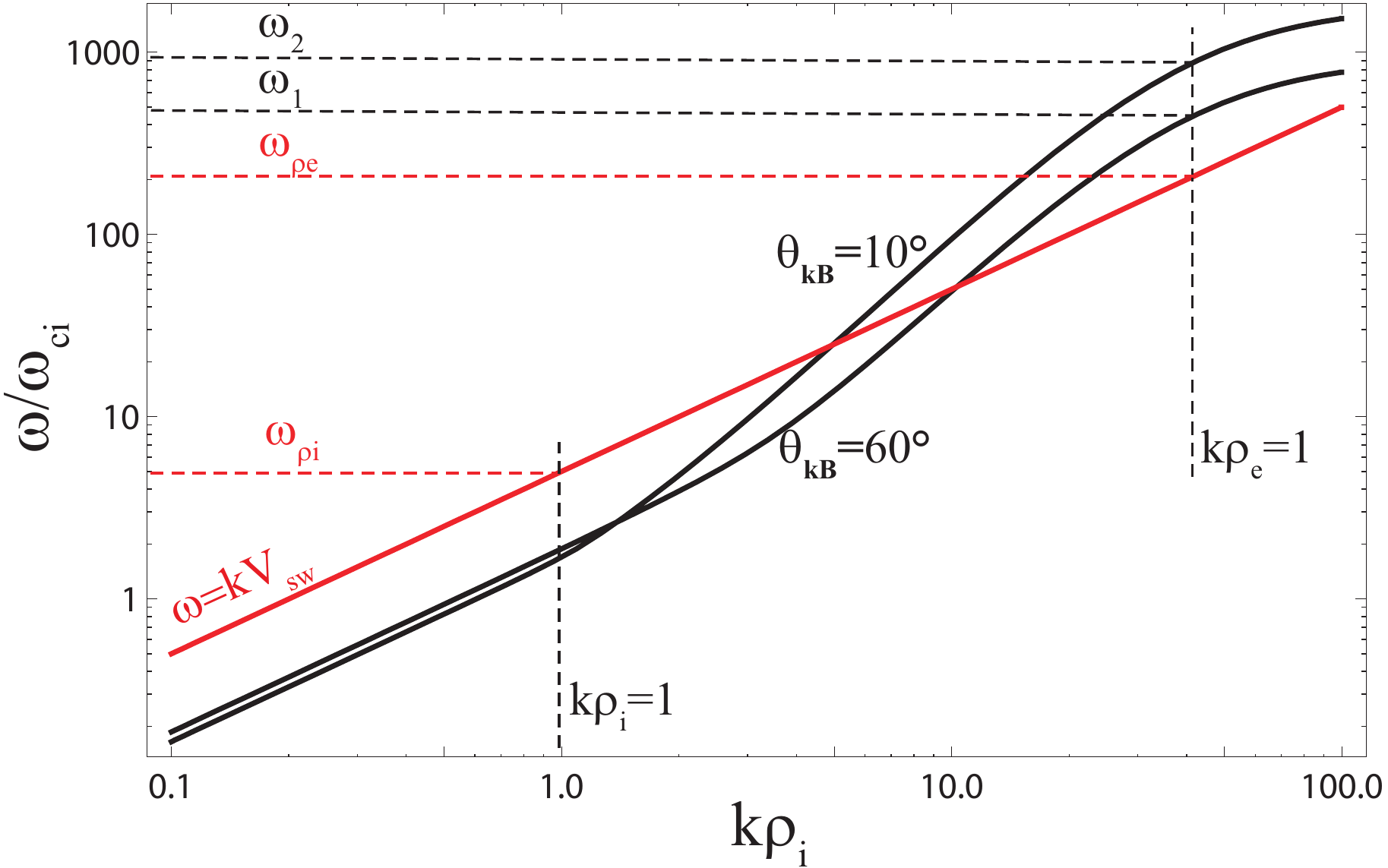}
\caption{Frequency signatures of the ion and electron gyroscales a frozen-in flow approximation (red) and two whistler modes at two different angles of propagation are used ($T_i=T_e$). The indicated frequencies are given in the text. \label{doppler}}
\end{figure}

When simultaneous measurement of $k_\parallel$ and $k_\perp$ is not possible, for instance when only single spacecraft data are available, another alternative, evoked briefly in \cite{sahraoui09}, exists to test which mode (whistler of KAW) is present in the data. As we will show below, this method is however limited to situations where turbulence is not confined to the very oblique angles studied here. Let us discuss it in more details here.  The general formula relating the frequency of the wave in the SW rest frame $\omega$ to the measured one onboard the spacecraft $\omega_{sc}$ is given by
\begin{equation}
\label{dopp}
\omega_{sc}=\omega+kV_{sw}\cos\theta_{\bf kV}
\end{equation}
where $V_{sw}$ is the SW speed forming an angle $\theta_{\bf kV}$ with the ${\bf k}$ vector.  If the KAW branch ($\omega<\omega_{ci}$) dominates  in the dispersive range down to the electron scale, then the Taylor frozen-in-flow assumption should be valid, meaning that equation~\ref{dopp} reduces to $\omega_{sc}\sim kV_{sw}\cos\theta_{\bf kV}$. Now, let us assume furthermore that two breakpoints occur in the turbulence energy spectra at the ion and the electron gyroscales $\rho_i$ and $\rho_e$~\citep{sahraoui09}. As shown in Fig.~\ref {doppler} these breakpoints should be observed in the spacecraft frame respectively at the Doppler-shifted frequencies $f_{\rho_i}=\omega_{\rho_i}/2\pi =V_{sw} \cos\theta_{\bf kV}/2\pi\rho_i$ and $f_{\rho_e}=\omega_{\rho_e}/2\pi =V_{sw} \cos\theta_{\bf kV}/2\pi\rho_e$. Assuming a similar angle $\theta_{\bf kV}$ at the ion and the electron scale, the ratio between the two frequencies should fulfill the relation 
\begin{equation}
\label{test}
\frac{f_{\rho_e}}{f_{\rho_i}} = \frac{\rho_i}{\rho_e} = \sqrt{\mu_{ei}\frac{T_i}{T_e}} \simeq 42\sqrt{\frac{T_i}{T_e}}
\end{equation}

From Fig.~\ref{doppler} one obtains the ratio $f_{\rho_e}/f_{\rho_i}\sim 42$ considering that $T_i=T_e$. In the data studied in \cite{sahraoui09} equation~\ref{test} has been accurately verified: with $T_i/T_e\sim 5$ the predicted ratio from equation~\ref{test} is $f_{\rho_e}/f_{\rho_i}\sim 95$, which was very close to the ratio between the frequencies corresponding to the observed electron breakpoint at $f_{\rho_e}\sim 40Hz$ and ion breakpoint at $f_{\rho_i}\sim 0.4Hz$. 

If however the relation~\ref{test} is not fulfilled by two observed breakpoints this would imply that the Taylor assumption fails, or equivalently, that a dispersive branch of the whistler type exists within the data. To show that let us now assume that the whistler branch is dominant in the dispersive range. In this case the Taylor hypothesis will naturally fail (since phase speeds comparable to $V_{sw}$ are present). At the very small scale, i.e., near $\rho_e$, one can even expect that the phase speed of the whistler mode, being linear in $k$ (since $\omega \propto k^2$), will be larger than $V_{sw}$ as can be seen in Fig.~\ref{doppler}. In this case equation~\ref{dopp} yields $\omega_{sc} \sim \omega$. This means that the relevant temporal scale will be given, not by ``Doppler-shifting'' $\rho_e$, but by the dispersion curve of the whistler mode in the plasma rest frame. From Fig.~\ref{doppler} one can see that the scale $\rho_e$ (=$d_e$) will yield a different breakpoint frequency depending on the angle $\theta_{\bf kB}$:  $\omega_1 \simeq 500 \omega_{ci}$ and $\omega_2 \simeq 1000 \omega_{ci}$ respectively for $\theta_{\bf kB}=60^\circ$ and $\theta_{\bf kB}=10^\circ$. With $f_{ci} \sim 0.5$Hz, the breakpoint at the electron scale should thus be observed onboard the spacecraft at the frequencies  $500$Hz and $250$Hz respectively. These frequencies are much higher than those reported in~\cite{sahraoui09}, which validates the conclusions of that work that turbulence was more consistent with the KAW modes (or quasi-stationary fluctuations) than with whistlers modes. At higher oblique angles ($\theta_{\bf kB}>60^\circ$) one can extrapolate from Fig.~\ref{doppler} that the phase speed of the whistler branch will become comparable or smaller than $V_{sw}$, therefore the Doppler-shift cannot be neglected in equation~\ref{dopp}. An accurate estimation of both terms in the right-hand side of equation~\ref{dopp} is then necessary. As mentioned above, this shows the limit of this simple method and emphasizes again the difficulty in distinguishing between the two branches, KAW and whistler, of the same mode at the very high oblique angles. The estimations given here can be tested on SW magnetic power spectra that can be measured by the high-time resolution Cluster search-coils (up to $225$Hz in the spacecraft frame of reference~\citep{sahraoui09,sahraoui10c}) to search for data intervals when parallel of moderate oblique whistler mode turbulence may exist (i.e. intervals not fulfilling equation~\ref{test}). However, very often, the level of the magnetic turbulence in the SW is not sufficiently high and can hit the sensitivity floor of the magnetometers at frequencies as low as $20Hz$. This limitation emphasizes the need in the future space missions for new search-coils that have higher levels of sensitivity in the dispersive and the dissipation ranges of SW turbulence. 

%--------------------------------------------------------------------------------------------
% New section 
%--------------------------------------------------------------------------------------------
\section{Conclusions}
Using the hot two-fluid and the Vlasov theories we studied key properties of the linear plasma modes under realistic SW conditions at 1AU, namely $\beta_i\gtrsim \beta_e \sim 1$, $T_i \gtrsim T_e$ and focused on high oblique angles of propagation $89^\circ \leq \theta_{\bf kB}<90^\circ$ as observed frequently from the Cluster multi-spacecraft data. We discussed the relevance of each of the main modes (the KAW, the whistler, the fast magnetosonic and the ion Bernstein) to carry the energy cascade below the ion gyroscale (the slow magnetosonic mode being heavily damped by kinetic effects). We showed in particular that the whistler branch develops as a continuation at high frequency of the classical (shear) Alfv\'en wave known in the MHD limit, and not of the fast magnetosonic mode. The Alfv\'en mode extends indeed at scales $k\rho_i\gtrsim1$ to frequencies either larger or smaller than $\omega_{ci}$, depending on the anisotropy $k_\parallel/ k_\perp$. The same mode is thus called whistler ($\omega>\omega_{ci}$) or KAW ($\omega<\omega_{ci}$) depending on the anisotropy ratio.

Unlike the two-fluid model, the Vlasov theory shows that the fast magnetosonic mode undergoes strong cyclotron damping at $\omega \sim \omega_{ci}$ and splits up into ion Bernstein modes at $\omega >\omega_{ci}$, in agreement with the analysis of the wave polarization. We showed also that ruling out the fast and the Bernstein modes from the cascade process in the dispersive range based on their electrostatic nature is incorrect, since the Alfv\'en-whistler modes are shown to be electrostatic as well at high oblique angles. We suggest rather that one must combine magnetic compressibility and damping rates to rule out those modes: both were shown to be highly compressible (i.e., $B_\parallel^2>>B_\perp^2$) in disagreement with typical SW observations of magnetic field data. In addition, those modes were shown to be heavily damped in the dispersive range as compared with the Alfv\'en-whistler mode. We showed finally that the modes on the whistler branch are more damped than those of the KAW branch, which makes the latter a more relevant candidate to carry the energy cascade down to electron scales. We also demonstrate a way to test the presence of either branch in SW spacecraft data and discussed the limitation of the existing spacecraft observations that need to be kept in mind in the planning of the future space missions.

Finally, it is important to keep in mind that the conclusions of this work may not be applicable to all SW observations because the present study is fully linear (one can question the relevance of linear theories to SW turbulence, see for instance~\cite{dmitruk09,howes09}) and restricted to the range of parameters discussed above. Moreover, the role that can be played by several plasma instabilities (both at ion and electron scales) as reported in several space observations~\cite{sahraoui06,hellinger06,bale09} is not addressed here.

\acknowledgments

\email{fouad.sahraoui@lpp.polytechnique.fr}

%\section{Appendix material}




\clearpage

\clearpage

\end{document}